\documentclass[12pt]{article}
\usepackage{amssymb,amsmath,amsthm,graphicx,ulem}

\usepackage{graphicx,subfigure}
\usepackage{epsfig}
\usepackage{amsmath}
\usepackage{amsfonts}
\usepackage{amssymb}
\usepackage[usenames]{color}
\usepackage[letterpaper,left=2.cm,right=2.cm,top=2.5cm,bottom=2.5cm]{geometry}
\usepackage{float}
\usepackage{xcolor}
\usepackage{hyperref}
\hypersetup{
    colorlinks = true,
    linkcolor = {black},
    urlcolor = {blue},
    citecolor = {red},
}
\usepackage{bm}
\usepackage[utf8]{inputenc}
\usepackage[english]{babel}
\usepackage{url}
\usepackage{authblk}
\usepackage{cite}

\def \dd {\mathrm{d}}
\def\d{\mathrm d}
\DeclareMathOperator{\ins}{\iota} 

\begin{document}
 
\allowdisplaybreaks
\normalem
\title{Covariant Noether charges for type IIB and 11-dimensional supergravities}

\author[1]{\'Oscar~J.~C.~Dias\footnote{ojcd1r13@soton.ac.uk}}
\author[1,2]{Gavin~S.~Hartnett\footnote{hartnett@rand.org}}
\author[3,4]{Jorge~E.~Santos\footnote{j.e.santos@damtp.cam.ac.uk}}

\affil[1]{STAG research centre and Mathematical Sciences, University of Southampton, UK}
\affil[2]{RAND Corporation, Santa Monica, CA 90401, USA}
\affil[3]{DAMTP, Centre for Mathematical Sciences, University of Cambridge, Wilberforce Road, Cambridge CB3 0WA, UK}
\affil[4]{Institute for Advanced Study, Princeton, NJ 08540, USA}

\date{}
\setcounter{Maxaffil}{0}
\renewcommand\Affilfont{\itshape\small}

\date{}
\maketitle 

\begin{abstract}
The covariant Noether charge formalism (also known as the covariant phase method) of Wald and collaborators, including its cohomological extension, is a manifestly covariant Hamiltonian formalism that, in principle, allows one to define and compute the energy, angular momenta, and chemical potentials of generic solutions of gravitational theories. However, it has been observed that for some supergravity solutions the variation of the Noether charge is not (at least manifestably) integrable, and as a result it is unclear whether there are well-defined thermodynamic charges for these solutions. In this work, we derive an expression for the variation of the covariant Noether charges for any solution of Type IIB 10-dimensional supergravity or 11-dimensional supergravity. Although this Noether quantity is not integrable in general, we show that for asymptotically scale-invariant solutions, it is. In particular, the asymptotic scale-invariance allows one to define an energy density and conjugate chemical potentials which obey a first law of thermodynamics and a Smarr relation. These two thermodynamic relations can then be shown to imply that the variation of the Noether charge is integrable, and accordingly the energy and other thermodynamic charges may be defined unambiguously. We explicitly demonstrate and illustrate our claim by computing the thermodynamic charges of two non-trivial supergravity solutions that were recently constructed: 1) the Polchinski-Strassler black brane that is dual to the deconfined phase of $\mathcal{N}=1^*$ theory, and 2) the CGLP black brane that asymptotes to the mass deformed Cveti\v{c}-Gibbons-L\"u-Pope (CGLP) solution.
\end{abstract}

\newpage
\tableofcontents
\baselineskip16pt


\section{Introduction}
The study of conservations laws in general relativity (and other gravitational theories), and the search for a robust and universal definition of gravitational energy are perhaps as old as general relativity itself. The first attempts to identify a local gravitational energy-momentum density began with the gravitational pseudo-tensor approaches, the most well known being perhaps the ones of Einstein and Landau-Lifshitz (see \cite{arnowitt1961coordinate, trautman1962conservation, brown1993quasilocal} for reviews). Yet these pseudo-tensors lead to coordinate-dependent expressions, and as a result they are not covariant and their geometrical interpretation is unclear\footnote{It is important to note that the non-tensorial nature of an object does not necessarily imply that it is meaningless. In particular, the Hamiltonian formalism endows the pseudo-tensors with physical meaning. Recall that in the Hamiltonian formulation of a gravitational theory, the Hamiltonian is decomposed as a sum of a spacetime volume contribution which vanishes on-shell (i.e, when the equations of motion are satisfied), and a boundary term whose value (when a choice of boundary condition is made) determines the thermodynamic charges. Ref.~\cite{chang1999pseudotensors} has shown that the known energy-momentum pseudo-tensors correspond to well identified Hamiltonian boundary terms once a specific boundary condition choice is made.}. This unsatisfactory property propelled subsequent studies where it became progressively clear that one cannot define a local and covariant energy-momentum density.\footnote{Essentially, this lack of a local definition of energy density follows from the Equivalence principle. Indeed, locally it is always possible to introduce a local inertial frame of reference associated to a free falling point-like particle. Such an observer does not feel any gravitational field and, accordingly, no gravitational energy density can be defined at spacetime points.} Rather, one must instead search for quasi-local or global definitions associated with extended regions in the spacetime, at least when the spacetime has an asymptotic symmetry. 

\subsection{Conserved Charges in General Relativity}
Historically, many different approaches for defining such conserved charges have been proposed. In order to provide context for the contributions of the present work, it will be useful to give a brief review of these various approaches and how they relate to one another. However, a full and comprehensive review of the history of conserved charges in gravitational theories is beyond the scope of this article, and the interested reader is encouraged consult the following excellent reviews: \cite{Ashtekar1980, frauendiener2004conformal, szabados2004quasi, hollands2005comparison, Compere:2007az, jaramillo2009mass,Chrusciel2010,papadimitriou2016lectures}.

Not surprisingly, the first considerations of conserved charges focused on asymptotically flat space-times (see \cite{Chrusciel2010} for a review). In the framework of the Hamiltonian formalism, the first major step forward in the task of defining conserved charges was given by the celebrated Arnowitt-Deser-Misner (ADM) definition of energy \cite{arnowitt1962dynamics}, which was then further complemented by the Regge-Teitelboim \cite{regge1974role} analysis\footnote{The proof of the positivity of the ADM energy, under the dominant energy condition for the matter energy-momentum tensor, is a keystone result of general relativity \cite{schoen1979proof,witten1981new}. Also note that for some spacetimes with certain isometries one can define a conserved Komar quantity that is the integral of the covariant derivative of a timelike Killing vector field over a $(d-2)-$surface \cite{komar1959covariant}. Sometimes, especially in the case of stationary asymptotically flat solutions, appropriate Komar integrals do match the corresponding ADM quantities \cite{Chrusciel2010}.}. The ADM mass and angular momentum are computed as integrals over spheres at spatial infinity, a calculation which in principle could be carried out, at least approximately, by a distant observer. Alternatively, one can use the Hamiltonian formalism with boundary at the null infinity; in this case we get the Bondi-Sachs mass and angular momentum \cite{bondi1962gravitational, sachs1962gravitational, wald2000general,Chrusciel2010}. And finally, using the Hamilton-Jacobi analysis of the gravitational action functional, which can be seen as a short-cut to the full Hamiltonian analysis,  quasi-local energy and angular momentum expressions (i.e. defined with respect to a bounded region of spacetime) were obtained by Brown and York \cite{brown1993quasilocal, brown2002action}. Reassuringly, these expressions converge to the ADM (or Bondi-Sachs) charges as spatial (or null infinity) is approached\footnote{Other useful quasi-local definitions of mass are the Misner-Sharp energy \cite{misner1964relativistic}, the Hawking mass \cite{hawking1968gravitational}, the Bartnik mass \cite{bartnik1989new}, the Kijowski$-$Liu-Yau energy \cite{kijowski1997simple, liu2003positivity}, the Wang-Yau mass \cite{wang2009quasilocal} among others (see \cite{szabados2004quasi, jaramillo2009mass, wang2009quasilocal,Chrusciel2010} for reviews on quasi-local masses).}. The original Hamiltonian analysis of \cite{arnowitt1962dynamics, regge1974role} was also extended to backgrounds with more generic asymptotics by Hawking and Horowitz \cite{hawking1996gravitational}\footnote{The expressions of \cite{hawking1996gravitational, hawking1996gravitationalb} reduce to the ADM expressions \cite{arnowitt1962dynamics} and Abbott-Deser expression \cite{abbott1982stability} when applied to the respective asymptotic backgrounds. The analysis of  \cite{brown1993quasilocal, brown2002action, hawking1996gravitational} requires a background subtraction procedure to get rid of divergences that arise due to the infinite spacetime volume. The same Hamiltonian and subtraction procedure can be used to compute the charges of spacetimes that are asymptotically the direct product of Minkowski spacetime and $\mathbb{R}^p$ (e.g. the black string or the black p-branes) \cite{harmark2004general, townsend2001first}.}. 

The first Hamiltonian analysis of energy for asymptotically  anti-de Sitter (AdS) spacetimes was provided by Abbott and Deser \cite{abbott1982stability} and further developed by Henneaux and Teitelboim \cite{henneaux1985asymptotically} and Chru\'sciel and Nagy \cite{Chrusciel:2000dd,Chrusciel:2001qr}. This analysis yields the conserved charges for pure AdS-Einstein gravity. However, it cannot be simply extended to AdS solutions which also have matter fields present, such as scalar fields with certain fall-offs. Such cases are particularly important in the context of the AdS/CFT correspondence. In this context, Skenderis and collaborators have developed the method of holographic renormalization \cite{henningson1998holographic, balasubramanian1999stress, de2001holographic, bianchi2002holographic, bianchi2001go, de2000holographic, kalkkinen2001holographic, martelli2003holographic, papadimitriou2004ads, elvang2016practical, skenderis2006kaluza, kanitscheider2008precision}, a formalism that computes the energy and other thermodynamic observables of gravitational systems that have a dual holographic gauge theory description\footnote{Within holographic renormalization, the gravitational conserved charges of an asymptotically (locally) AdS solution is computed from the energy momentum tensor that one obtains when one varies the on-shell gravitational action w.r.t. the boundary metric. In spirit, this approach is thus similar to the traditional Hamiltonian analysis of, for example  \cite{brown1993quasilocal,hawking1996gravitational}. However, this computation generically yields a divergent answer due to the infinite volume of the background. To extract the physical finite answer, it is common to use a background subtraction method (see e.g. \cite{brown1993quasilocal, hawking1996gravitational}); however an appropriate reference spacetime is not guaranteed to exist in general. Holographic renormalization effectively implements this background subtraction procedure in AdS in a systematic way using boundary counterterms. This is a procedure that follows naturally from the AdS/CFT duality since this is exactly the approach used in  the QFT process of renormalization to remove unphysical divergencies.}. A Hamilton-Jacobi approach to the method of holographic renormalization, whereby the radial coordinate plays the role of time was also developed  \cite{de2000holographic, kalkkinen2001holographic, martelli2003holographic, papadimitriou2004ads, papadimitriou2005thermodynamics,elvang2016practical}\footnote{Positivity energy theorems for asymptotically AdS and asymptotically locally AdS spacetimes has been proven in \cite{gibbons1983stability, cheng2005positivity} using the Witten-Nester spinorial energy \cite{witten1981new, nester1981new}.}. It should be emphasized that the holographic renormalization charges agree with previous definitions of conserved charges \cite{abbott1982stability, henneaux1985asymptotically, ashtekar1984asymptotically, hawking1996gravitational, ashtekar2000asymptotically,papadimitriou2005thermodynamics} when the latter are applicable, i.e. when the spacetime is conformal to the AdS solution and one considers the energy relative to AdS \cite{papadimitriou2005thermodynamics,hollands2005comparison}. Holographic renormalization also may be used for arbitrary asymptotically locally AdS spacetimes. Additionally, it may be shown that the holographic renormalization charges may be interpreted as Noether charges associated to asymptotic symmetries of the gravitational solution \cite{papadimitriou2005thermodynamics}, and that these agree with the conserved charges computed using the covariant Noether charge approach, also called the covariant phase space approach \cite{wald1993black, iyer1994some, iyer1995comparison, wald2000general,papadimitriou2005thermodynamics}.

Another important class of conserved charges are those provided by the covariant Noether charge formalism of Wald and collaborators \cite{wald1993black, iyer1994some, iyer1995comparison, wald2000general,papadimitriou2005thermodynamics}, and is a systematic improvement of the covariant phase or covariant sympletic method initially discussed in \cite{ashtekar1991covariant, nester1991covariant}. It also generalizes and incorporates many of the previous Hamiltonian approaches. Indeed, the covariant Noether charge formalism could equally well be called the manifestly Lorentz-covariant Hamiltonian formalism since it avoids the $(D-1) + 1$ decomposition between spatial coordinates and the time coordinate that the ADM formalism requires.\footnote{Of course, with the traditional ADM and Hamiltonian approaches we can always check {\it \`a posteriori} that the conserved quantities obtained in the $(D-1) + 1$ form are actually Lorentz-covariant. The first manifestly covariant Hamiltonian analysis has been developed by Abbott and Deser \cite{abbott1982stability}, which was also the first satisfactory approach to study charges in asymptotically AdS spacetimes.}

Lastly, two other major classes of methods which deserve mention are conformal methods \cite{penrose1963asymptotic, geroch1977asymptotic, ashtekar1978unified, christodoulou2014global} and spinorial methods \cite{witten1981new, nester1981new, gibbons1983positive, gibbons1983stability, cheng2005positivity,Chrusciel2010}. A particularly relevant sub-class of conformal methods are those which which exploit the asymptotic structure of spacetimes to compute the energy. This approach is best reviewed in \cite{Ashtekar1980} and it has been applied to asymptotic flat backgrounds \cite{penrose1963asymptotic, geroch1977asymptotic, ashtekar1978unified, christodoulou2014global} as well to asymptotically AdS spacetimes \cite{ashtekar1984asymptotically, ashtekar2000asymptotically}. Spinorial methods have been used to prove the positivity of energy both in asymptotically flat \cite{witten1981new} and asymptotically AdS \cite{gibbons1983positive,gibbons1983stability}, with our without black hole horizons (see review \cite{Chrusciel2010}). In \cite{cheng2005positivity} attempts have been made to extend positivity of energy to general asymptotically locally AdS spacetimes.

\subsection{Integrability and Covariant Noether Charges}
In the present study, we are particularly interested in using the covariant Noether charge formalism of Wald and collaborators \cite{wald1993black, iyer1994some, iyer1995comparison, wald2000general,papadimitriou2005thermodynamics} and its cohomological formalism extension \cite{Anderson:1996sc,Barnich:2001jy,Barnich:2003xg,Barnich:2007bf,Compere:2007vx,Chow:2013gba,Compere:2007az} to compute the energy and thermodynamic potentials of gravitational theories.

An aspect of the covariant Noether charge formalism that is not completely satisfactory is the fact that it still has ambiguities associated to the fact that we can always add a total divergence to the action. Using Stokes' theorem this amounts to have a freedom in adding a boundary term to the action and Hamiltonian that can still change the definition of charges.
Lagrangian methods based on cohomological techniques attempt to eliminate this feature. That is, the cohomological formalism is the most recent attempt to define conserved charges in gravitational theories in a more mathematically rigorous way \cite{Anderson:1996sc,Barnich:2001jy,Barnich:2003xg,Barnich:2007bf,Compere:2007vx,Chow:2013gba} (see \cite{Compere:2007az} for a detailed review). Essentially, when the symmetry associated to the charge (energy, angular momentum, ...) under consideration is exact, this method incorporates and agrees with the covariant Noether charge (phase space) formalism of \cite{Wald:1993nt,Iyer:1994ys,Iyer:1995kg,Wald:1999wa}. However, in the cohomological formalism, surface charges are computed integrating surface charge one-forms. The latter are constructed directly from Euler-Lagrange derivatives of the gravitational Lagrangian, which do not depend on total divergence terms. Therefore, the cohomological approach formally avoids the above ambiguities of the Noether charge formalism (in practice, it is important to reinforce that the cohomological method yields the same charges as the covariant Noether charges if in the latter formalism we just consider the boundary terms that arise naturally when we use integration by parts to eliminate derivatives of the field that is being  varied to get the equations of motion). Additionally, but not less important, the cohomological method extends the covariant phase space method in the sense that it can also be used to define conserved charges that are associated to asymptotic (as oppose to exact) symmetries (in this manuscript we shall not consider such cases). Henceforth, in a slight abuse of language, when we refer to the Noether charge formalism we will often be implicitly referring to the cohomological extension of the standard Noether approach.

It has been observed and highlighted that the covariant Noether charge and cohomological formalisms can have a limitation or undesirable feature \cite{Wald:1999wa,Compere:2007az,Chow:2013gba}. Namely, there are systems where the lack of {\it manifest} integrability of the first order variation of the charges seems to impede a definition of the charges  \cite{Wald:1999wa,Compere:2007az,Chow:2013gba}.
Indeed, these two formalisms yield an expression for the {\it variation} of a Noether charge along the moduli space of parameters that parametrize a given family of solutions. In order to get the actual conserved charge of this family of solutions, one must then integrate this charge variation along a path starting from a reference solution. For one-parameter systems this is not a problem. However, in general $-$ for systems described two or more parameters $-$ we are often not guaranteed that such a conserved charge is well-defined. In the optimal scenario, certain integrability conditions are satisfied and the existence of the desired well-defined conserved charge for the solution at hand is guaranteed \cite{Wald:1999wa,Compere:2007az,Chow:2013gba}. Notably, this is the case for the Kerr solution. {\it However}, supergravity solutions have already been found where  integrability conditions are not obeyed and thus it was declared that energy is simply not defined for such a system \cite{Chow:2013gba}. In this manuscript, we will add a few more examples $-$ the Polchinski-Strassler black brane \cite{Bena:2018vtu} and the CGLP black brane \cite{Dias:2017opt}.

The main aim of the present work is to observe that even when the Noether charge variations are not manifestly integrable, asymptotic scale invariance allows two thermodynamic relations to be derived which in turn imply that that the Noether variation is in fact integrable. The two relations are: a first law of thermodynamics and a Smarr relation. Asymptotic scale-invariance is a common property for gravitational solutions in the context of the gauge/gravity correspondence. In particular it holds for solutions which are asymptotically a direct product spacetime $AdS_p\times X^n$ where $X^n$ is a compact manifold. We illustrate our claim explicitly by considering three non-trivial examples in IIB \cite{Bena:2018vtu} and 11-dimensional supergravity \cite{Dias:2017opt,Costa:2014wya}.

The layout of this article is as follows. In Sec.~\ref{sec:IIB} we start by using the covariant Noether charge formalism of \cite{wald1993black, iyer1994some, iyer1995comparison, wald2000general} to define the variation of the Noether conserved charge of any type IIB supergravity solution. Remarkably, this expression has never appeared in the literature, as far as we are aware. As an application of this expression to a manifestly integrable system, in Sec.~\ref{sec:lumpyLoc} we compute the energy of the IIB lumpy and localized black holes of \cite{dias2015lumpy, dias2016localized}. Not surprisingly, we will find that this Noether energy agrees with the energy computed originally using Kaluza-Klein holographic renormalization. Next, in Sec.~\ref{sec:PSbh} we  discuss the energy of the Polchinski-Strassler black brane recently constructed in \cite{Bena:2018vtu}. This is an asymptotically AdS$_5\times S^5$ solution of IIB supergravity where all the bosonic fields are non-trivial and that corresponds to the high-temperature phase of the gravitational-dual of the Polchinski-Strassler solution \cite{polchinski2000string}. In this case, the variation of the Noether charge does not lead to an obviously integrable expression. By exploiting the asymptotic scale-invariance, we derive two additional relations, a thermodynamic first law and a Smarr relation. When these relations are satisfied, the Noether variation becomes explicitly integrable, and an energy and chemical potential may then be defined unambiguously. We note that for this example, current holographic renormalization techniques cannot be used to compute the desired charges. 

In Sec.~\ref{sec:Mtheory} we consider the low-energy classical limit of M-theory. Again, we will start by using the covariant Noether charge formalism of \cite{wald1993black, iyer1994some, iyer1995comparison, wald2000general} to compute the variation of the Noether conserved charge of a general solution to 11-dimensional supergravity, an expression that is missing in the literature. Then, in Sec.~\ref{subsec:CGLPbh} we will consider the thermodynamics of the CGLP black brane recently constructed in \cite{Dias:2017opt}. This is an asymptotically $AdS_4\times V_{5,2}$ solution where $V_{5,2}$ is the 7-dimensional Stiefel manifold. It is the simplest finite-temperature solution that asymptotes to the Cveti\v{c}, Gibbons, L\"u, and Pope (CGLP) solution \cite{cvetivc2003ricci}, which has a magnetic $G_{(4)}$-flux background whose asymptotic decay describes a mass deformation of the corresponding dual $CFT_3$. As before, we will find that the associated Noether variation charge does not satisfy the required integrability conditions which could suggest that an energy might not be well defined for this system. However, we will again find that asymptotic scale-invariance may be used to show that the Noether variation is explicitly integrable, and we are thus able to derive the desired thermodynamic charges.
\section{10-Dimensional IIB Supergravity \label{sec:IIB}}
The massless bosonic fields of  type IIB supergravity are the metric tensor $g_{ab}$, the  dilaton $\Phi$,  the axion $C$, the NS-NS antisymmetric  2-tensor $B_{(2)}$,  the R-R 2-form potential $C_{(2)}$, and the R-R 4-form field $C_{(4)}$ with self-dual 5-form field strength (their fermionic superpartners are a complex Weyl gravitino and a complex Weyl dilatino). We can combine the dilaton and axion scalars into a complex field $B$, and  define a complex 2-form $A_{(2)}$ that is a combination of the NS-NS field $B_{(2)}$ and the R-R potential $C_{(2)}$ \cite{schwarz1983covariant,grana2002gauge}:
\begin{eqnarray}
&& B \equiv \frac{1+i\tau}{1-i\tau}\,, \qquad \tau \equiv C + i \,e^{-\Phi}\,; \nonumber \\
&& A_{(2)} \equiv \frac{g_s}{\kappa_{10}}\, \left( B_{(2)} + i \,C_{(2)} \right),
\label{IIBfieldCombinations}
\end{eqnarray}
where $g_s=e^{\Phi_\infty}$ is the string coupling and $\kappa_{10}$ is related to Newton's constant by $2\kappa_{10}= 16\pi G_{10}$ (which can also be written in terms of $g_s$ and the string length $\ell_s$ as $2\kappa_{10}=(2 \pi )^7 g_s^2 \ell_s^8$ ).

Taking the real fields $g,C_{(4)}$ and the complex fields $B,A_{(2)}$ as fundamental fields, the bosonic action of type IIB supergravity, in the Einstein frame, reads \cite{schwarz1983covariant,grana2002gauge}
\begin{equation}
S_{\text{IIB}} = \frac{1}{2 \kappa_{10}^2} \int \bigg( \star\, \mathcal{R} -2 P\wedge \star\, \bar{P}- \frac{1}{2} \, G_{(3)} \wedge \star\, \bar{G}_{(3)} -4 F_{(5)}\wedge \star\,F_{(5)}- i \, C_{4} \wedge G_{(3)} \wedge \bar{G}_{(3)} \bigg)\,,
\label{IIBsugraAction}
\end{equation}
supplemented by the self-duality condition 
\begin{equation}\label{IIB:selfdual}
\star F_{(5)} = F_{(5)}\,.
\end{equation}
In this action, $\mathcal{R}$ is the Ricci volume form, $\mathcal{R}=R\,\star \mathbb{I}=R\,\mathrm{vol}_{10}$,  the bar over a quantity denotes complex conjugation, and we have defined
\begin{eqnarray}\label{IIBsugraAction2}
&& f = \frac{1}{\sqrt{1-|B|^2}}\,, \qquad P = f^2\,\mathrm{d} B\,, \qquad Q = f^2\,\mathrm{Im}(B\,\mathrm{d} B)\,, \nonumber \\
&& G_{(3)} = f\,(F_{(3)}-B\bar{F}_{(3)})\,, \qquad  F_{(3)} = \mathrm{d}A_{(2)}\,,
\nonumber \\
&&F_{(5)} = \mathrm{d}C_{(4)}-\frac{1}{8}\mathrm{Im}(A_{(2)}\wedge \bar{F}_{(3)}) =\mathrm{d}C_{(4)}+\frac{i}{16}(A_{(2)}\wedge \bar{F}_{(3)}-\bar{A}_{(2)}\wedge F_{(3)})\,,  
\end{eqnarray}
where $Q$ is introduced here for later use.

\subsection{Variation of the Noether Charge}
The Noether charge formalism proceeds by considering the variation of the action. As usual, the first order variation of the action \eqref{IIBsugraAction} leads to the classical equations of motion. The variation will include boundary terms which do not affect the equations of motion, however these are fundamental for computing conserved charges. Consider first the variation with respect to the graviton, which yields:
\begin{equation}\label{IIB:varG}
 \delta S_{\text{IIB}}\big |_g=\frac{1}{2 \kappa_{10}^2} \int \mathrm{d}^{10}x\sqrt{-g}\left[ R_{ab}-\frac{1}{2} \,R g_{ab}- T^{(1)}_{ab}-T^{(3)}_{ab}-T^{(5)}_{ab}\right] \delta g^{ab} + \delta S_{g}^{(\text{GH})} \,,
\end{equation}
with
\begin{eqnarray}\label{IIB:varG2}
&& T^{(1)}_{ab}=P_a\bar{P}_b+P_b\bar{P}_a-g_{ab}P_c \bar{P}^c\,,
\nonumber \\
&& T^{(3)}_{ab}=\frac{1}{8}\left[(G_{(3)})_a^{\phantom{a}cd}(\bar{G}_{(3)})_{bcd}+(G_{(3)})_b^{\phantom{a}cd}(\bar{G}_{(3)})_{acd}-\frac{1}{3}g_{ab} (G_{(3)})^{abc}(\bar{G}_{(3)})_{abc}\right]\,,\nonumber \\
&& T^{(5)}_{ab}=\frac{1}{6}(F_{(5)})_{a}^{\phantom{a}cdef}(F_{(5)})_{bcdef}\,.
\end{eqnarray}
$\delta S_{g}^{(\text{GH})}$ is a total divergence contribution, namely the well-known Gibbons-Hawking boundary term 
\begin{equation}\label{IIB:bdryG0}
 \delta S_{g}^{(\text{GH})}= \frac{1}{2 \kappa_{10}^2} \int \mathrm{d}^{10}x \sqrt{-g} \,g^{ab}\,\delta R_{ab}=\frac{1}{2 \kappa_{10}^2} \int \mathrm{d}^{10}x \sqrt{-g} \, \nabla^a  \left( \nabla^b \delta g_{ab}-\nabla_a \delta g\right).  
\end{equation}

It will be convenient to encode the information of this Gibbons-Hawking boundary term into a 9-form:
\begin{equation}\label{IIB:bdryG}
\theta(g,\delta g)=\frac{1}{2 \kappa_{10}^2}  \star\left[ \left( \nabla^b \delta g_{ab}-\nabla_a \delta g\right) \dd x^a \right].  
\end{equation}

Next, take the variation with respect to the axion/dilaton complex scalar $B$: 
\begin{eqnarray}\label{IIB:varB}
 \delta S_{\text{IIB}}\big |_{B,\bar{B}} &=&  \frac{1}{2 \kappa_{10}^2} \int \delta \big |_{B,\bar{B}}\left[ -2 P\wedge \star\, \bar{P}- \frac{1}{2} \, G_{(3)} \wedge \star\, \bar{G}_{(3)} - i \, C_{(4)} \wedge G_{(3)} \wedge \bar{G}_{(3)} \right] \nonumber \\
 &=& \frac{1}{2 \kappa_{10}^2} \int 2 f^2  \bigg[ \left( \mathrm{d} \star \bar{P} +2 i \,\bar{Q}\wedge \star \bar{P} +\frac{1}{4} \bar{G}_{(3)}\wedge \star \,\bar{G}_{(3)} \right) \delta B  + {\rm c.c.}  \bigg]+   \theta\left( B,\bar{B},\delta B,\delta\bar{B}\right), \nonumber \\
\end{eqnarray}
where $Q$ is defined in \eqref{IIBsugraAction2}, ${\rm c.c.} $ stands for complex conjugate, and the boundary term is 
\begin{equation}\label{IIB:bdryB}
 \theta\left( B,\bar{B},\delta B,\delta\bar{B}\right)=-\frac{1}{\kappa_{10}^2}f^2\left( \star\,\bar{P}\wedge\delta B+\star\,P\wedge\delta \bar{B}\right).
\end{equation}

Consider now the variation of the action with respect to the complex scalar $A_{(2)}$. A long computation yields:\footnote{Recall that $F_{(5)}$ is self-dual, $\star F_{(5)}=F_{(5)}$.}
\begin{eqnarray}\label{IIB:varA2}
 \delta S_{\text{IIB}}\big |_{A_{(2)},\bar{A}_{(2)}} \hspace{-0.3cm}&=& \hspace{-0.2cm}  \frac{1}{2 \kappa_{10}^2} \int \delta \big |_{A_{(2)},\bar{A}_{(2)}} \left[ - \frac{1}{2} \, G_{(3)} \wedge \star\, \bar{G}_{(3)} -4 F_{(5)}\wedge \star\,F_{(5)}- i \, C_{(4)} \wedge G_{(3)} \wedge \bar{G}_{(3)}  \right] \nonumber \\
 &=& \hspace{-0.2cm} \frac{1}{2 \kappa_{10}^2} \int \frac{1}{2f}  \bigg[ \left( \mathrm{d} \star \bar{G}_{(3)} + i \,\bar{Q}\wedge \star \,\bar{G}_{(3)} -\bar{P}\wedge\star\,G_{(3)}-4\,i\,\bar{G}_{(3)}\wedge \star \,F_{(5)} \right) \wedge \delta A_{(2)} + {\rm c.c.}   \bigg] \nonumber \\
&& \hspace{-0.2cm} + \theta\left(A_{(2)},\bar{A}_{(2)},\delta A_{(2)},\delta\bar{A}_{(2)}\right), 
\end{eqnarray}
where the boundary term is 
\begin{equation}
\label{IIB:bdryA2}
\theta\left(A_{(2)},\bar{A}_{(2)},\delta A_{(2)},\delta\bar{A}_{(2)}\right) = \frac{1}{4\kappa_{10}^2} \Big[ i \star F_{(5)} \wedge\bar{A}_{(2)}- f\left( \star\, \bar{G}_{(3)}-\bar{B}\star G_{(3)}\right) - 2i \,C_{(4)}\wedge\bar{F}_{(3)} \Big]\wedge\delta A_{(2)}+ {\rm c.c.}   
\end{equation}

Finally, consider the variation of the action with respect to the real R-R field $C_{(4)}$ which gives:
\begin{eqnarray}\label{IIB:varA4}
 \delta S_{\text{IIB}}\big |_{C_{(4)}} &=&  \frac{1}{2 \kappa_{10}^2} \int \delta \big |_{C_{(4)}}\left[ -4 F_{(5)}\wedge \star\,F_{(5)} - i \, C_{(4)} \wedge G_{(3)} \wedge \bar{G}_{(3)} \right] \nonumber \\
 &=& \frac{4}{\kappa_{10}^2} \int \left( \mathrm{d} \star F_{(5)}-\frac{i}{8}G_{(3)}\wedge\bar{G}_{(3)} \right) \wedge \delta C_{(4)}  
 +   \theta\left( C_{(4)}, \delta C_{(4)} \right)
\end{eqnarray}
with the boundary term given by 
\begin{equation}\label{IIB:bdryA4}
 \theta\left( C_{(4)}, \delta C_{(4)} \right)=-\frac{4}{\kappa_{10}^2} \star F_{(5)}\wedge \delta C_{(4)}\,.
\end{equation}

Requiring that the variations \eqref{IIB:varG}, \eqref{IIB:varB}, \eqref{IIB:varA2}, \eqref{IIB:varA4} vanish leads to the equations of motion of type IIB supergravity:
\begin{subequations}
\label{eqs:IIB}
\begin{align}
&R_{ab} = P_a\bar{P}_b+P_b\bar{P}_a  + \frac{1}{6}(F_{(5)})_{a}^{\phantom{a}cdef}(F_{(5)})_{bcdef} \nonumber \\
& \qquad \qquad + \frac{1}{8}\left[(G_{(3)})_a^{\phantom{a}cd}(\bar{G}_{(3)})_{bcd}+(G_{(3)})_b^{\phantom{a}cd}(\bar{G}_{(3)})_{acd}-\frac{1}{6}g_{ab} (G_{(3)})^{abc}(\bar{G}_{(3)})_{abc}\right] \,,
\label{eq:einstein}
\\
& \mathrm{d} \star P -2 i \,Q\wedge \star \,P +\frac{1}{4} G_{(3)}\wedge \star\, G_{(3)} = 0\,,
\\
&\mathrm{d}\star G_{(3)}-i \,Q\wedge \star \,G_{(3)}-P\wedge \star \,\bar{G}_{(3)}+4i \,G_{(3)}\wedge \star F_{(5)}=0\,,
\label{eq:g3}
\\
&\mathrm{d} \star F_{(5)}-\frac{i}{8}G_{(3)}\wedge\bar{G}_{(3)}=0\,, \label{eq:f5}
\end{align}%
\end{subequations}
where we contracted  \eqref{IIB:varG}-\eqref{IIB:varG2} with the inverse metric to get the Ricci scalar $R$ and then inserted this quantity back into  \eqref{IIB:varG} to get the trace reversed equation of motion for the graviton \eqref{eq:einstein}.

It follows from \eqref{eq:einstein} that the on-shell Ricci volume form is $\star\mathcal{R}=2 P\wedge\star \bar{P}+\frac{1}{4}G_{(3)}\wedge\star\bar{G}_{(3)}$ . Moreover, the self-duality condition \eqref{IIB:selfdual} implies that on-shell $F_{(5)}\wedge \star F_{(5)} =F_{(5)}\wedge F_{(5)} =0$. Using these relations on \eqref{IIBsugraAction} one finds that the on-shell 10-form Lagrangian reads
\begin{equation}\label{IIB:onshellL}
\mathcal{L}\big |_{on-shell} = -\frac{1}{2 \kappa_{10}^2} \left( \frac{1}{4} G_{(3)}\wedge \star \,\bar{G}_{(3)} +i\, C_{(4)} \wedge G_{(3)} \wedge \bar{G}_{(3)} \right).
\end{equation}

Given a diffeomorphism vector generator $\xi$, we can construct the associated sympletic Noether current 9-form \cite{wald1993black, iyer1994some, wald2000general}:
 \begin{align}\label{IIB:current}
J =\Theta\left(g,\mathcal{L}_\xi g\right) &+ \Theta\left(P,\bar{P},\mathcal{L}_\xi P,\mathcal{L}_\xi \bar{P}\right)+\Theta\left(A_{(2)},\bar{A}_{(2)},\mathcal{L}_\xi A_{(2)},\mathcal{L}_\xi \bar{A}_{(2)} \right) \nonumber \\
&+ \Theta(C_{(4)},\mathcal{L}_\xi C_{(4)})- \ins_\xi \mathcal{L}\big |_{on-shell} \,,
\end{align}
where $\Theta(\phi_i,\mathcal{L}_\xi \phi_i)\equiv \theta(\phi_i,\mathcal{L}_\xi \phi_i)$ for $\phi_i = \{g,P,\bar{P},A_{(2)},\bar{A}_{(2)},C_{(4)} \}$, i.e. we make the replacements $\delta \phi_i\to \mathcal{L}_\xi \phi_i$ on the boundary terms \eqref{IIB:bdryG}, \eqref{IIB:bdryB}, \eqref{IIB:bdryA2} and \eqref{IIB:bdryA4}. Here, $\mathcal{L}_\xi \phi_i$ is the  Lie derivative of the field $\phi_i$ along the diffeomorphism generator $\xi$. Also, $ \ins_\xi \mathcal{L}\big |_{on-shell} $ is the interior product  of $\xi$ with the 10-form \eqref{IIB:onshellL}. 

It can be shown that $\dd J=- E_i \,\mathcal{L}_\xi \phi_i$, where $E_i$ stands for the equations of motion (summation convention holds here) \cite{wald1993black, iyer1994some, wald2000general}. Therefore, on-shell ($E_i=0$) the current is closed, i.e. $\dd J=0$ for all $\xi$. It follows that there is a Noether charge 8-form $\widetilde{Q}_\xi$ locally constructed from $\{\xi, \phi_i\}$, such that on-shell one has $J=\dd \widetilde{Q}$, since in these conditions $\dd J=\dd^2\widetilde{Q}_\xi=0$ \cite{wald1993black, iyer1994some, wald2000general}.\footnote{$\widetilde{Q}_\xi$ is defined uniquely up to the addition of a closed form $\dd \chi$.} 

To evaluate \eqref{IIB:current} it is useful to recall the definition of Lie derivative of a scalar\footnote{This is Cartan's formula for $p=0$.} and of a torsion-free metric tensor (with a Levi-Civita connection), as well as Cartan's formula for the Lie derivative of a $p$-form $A_{(p)}$:  
 \begin{eqnarray}\label{Cartan}
&& \mathcal{L}_\xi B=\xi^a\nabla_a B =\ins_\xi d B\,, \nonumber\\
&& \mathcal{L}_\xi g_{ab}=2\nabla_{(a} \xi_{b)}\,,\nonumber\\
&&  \mathcal{L}_\xi A_{(p)} = \dd \ins_\xi A_{(p)}+ \ins_\xi \dd A_{(p)}\,.
\end{eqnarray}

Using these relations, the identity $G_{(3)}\wedge \bar{G}_{(3)}=F_{(3)}\wedge \bar{F}_{(3)}$, the equations of motion \eqref{eqs:IIB}, and the $p$-form identities listed in the appendix in \eqref{pformIdentities}, one finds:\footnote{To get $\Theta\left(g,\mathcal{L}_\xi g\right)$ we further use the commutator relation $\left[\nabla_a,\nabla_b\right]\xi_c=R_{abcd}\xi^d$}   
 \begin{eqnarray}\label{IIB:ThetaG}
 \Theta\left(g,\mathcal{L}_\xi g\right)&=&\star\Big( \frac{1}{\kappa_{10}^2}\left[\nabla^b \nabla_{(a}\xi_{b)} -g^{bc}\nabla_a\nabla_{(b}\nabla_{c)}\right]\dd x^a \Big)\nonumber\\
 &=&\frac{1}{9!}\left[ \frac{1}{\kappa_{10}^2} \varepsilon_{a_1\cdots a_9 a}\left( \nabla_b \nabla^{[b}\xi^{a]}+R^a_{\:\:b}\xi^b\right)\right]\dd x^{a_1}\wedge \cdots \wedge \dd x^{a_9}\,, 
\end{eqnarray}
 \begin{eqnarray}\label{IIB:ThetaB}
\Theta\left(B,\bar{B},\mathcal{L}_\xi B,\mathcal{L}_\xi \bar{B}\right)&=&-\frac{1}{\kappa_{10}^2}\,f^2 \left( \star\bar{P}\wedge \ins_\xi dB +\star P\wedge \ins_\xi d\bar{B}\right)\nonumber\\
&=&\frac{1}{9!}\left[ \frac{1}{\kappa_{10}^2} \varepsilon_{a_1\cdots a_9 a}\left( -\bar{P}^aP_b-P^a \bar{P}_b\right)\xi^b \right]\dd x^{a_1}\wedge \cdots \wedge \dd x^{a_9}\,, 
\end{eqnarray}
 \begin{eqnarray}\label{IIB:ThetaA4}
   \Theta\left( C_{(4)}, \mathcal{L}_\xi C_{(4)} \right) &=&-\frac{4}{\kappa_{10}^2} \star F_{(5)}\wedge \left[ \ins_\xi \dd C_{(4)} + \dd \left(\ins_\xi C_{(4)} \right) \right] \nonumber\\
&=& \frac{1}{9!} \left[ \frac{1}{\kappa_{10}^2} \varepsilon_{a_1\cdots a_9 a}\left( 
-\frac{1}{6} (F_{(5)})^{a c_1\cdots c_{(4)}}(F_{(5)})_{b c_1\cdots c_4}\right)\xi^b \right]\dd x^{a_1}\wedge \cdots \wedge \dd x^{a_9}\nonumber\\
&& +\frac{1}{4\kappa_{10}^2} \big( -i\star F_{(5)}\wedge \bar{F}_{(3)}\wedge\ins_\xi A_{(2)} + i\star F_{(5)}\wedge A_{(2)} \wedge\ins_\xi  \bar{F}_{(3)} + {\rm c.c.}  \big) \nonumber\\
&&+ \frac{4}{\kappa_{10}^2} \dd\left( \star F_{(5)} \wedge \ins_\xi C_{(4)} \right)-\frac{4}{\kappa_{10}^2} \dd \star F_{(5)} \wedge \ins_\xi C_{(4)} \,, 
\end{eqnarray}
 \begin{eqnarray}\label{IIB:ThetaA2}
&& \hspace{-0.6cm}    \Theta\left( A_{(2)},\bar{A}_{(2)}, \mathcal{L}_\xi A_{(2)},\mathcal{L}_\xi \bar{A}_{(2)}  \right) \nonumber\\
&& \hspace{0.5cm} = \frac{1}{4\kappa_{10}^2}\Big[ i \star F_{(5)} \wedge\bar{A}_{(2)} -f\left( \star\, \bar{G}_{(3)}-\bar{B}\star G_{(3)}\right) 
   -2i \,C_{(4)}\wedge\bar{F}_{(3)}\Big]\wedge\Big[ \ins_\xi F_{(3)} + \dd \left(\ins_\xi A_{(2)} \right) \Big]  +{\rm c.c.}   \nonumber\\
&& \hspace{0.2cm}  = \frac{1}{9!} \left[ \frac{1}{\kappa_{10}^2} \varepsilon_{a_1\cdots a_9 a}\left( 
-\frac{1}{8} (G_{(3)})^{acd}(\bar{G}_{(3)})_{bcd}-\frac{1}{8}(\bar{G}_{(3)})^{acd}(G_{(3)})_{bcd} \right)\xi^b \right]\dd x^{a_1}\wedge \cdots \wedge \dd x^{a_9}\nonumber\\
&& \hspace{0.7cm}+ \frac{1}{4 \kappa_{10}^2} \Big[ i\star F_{(5)} \wedge \bar{F}_{(3)}\wedge \ins_\xi A_{(2)} +\left( i\star F_{(5)} \wedge \bar{A}_{(2)} -2i \,C_{(4)}\wedge \bar{F}_{(3)} \right)\wedge \ins_\xi F_{(3)} +{\rm c.c.}   \Big] \nonumber\\
&& \hspace{0.7cm} + \dd \left[ \frac{1}{4\kappa_{10}^2}\left[ f\left( \star\, \bar{G}_{(3)}-\bar{B}\star G_{(3)}\right) - i \star F_{(5)} \wedge\bar{A}_{(2)} +2i \,C_{(4)}\wedge\bar{F}_{(3)} \right]\wedge \ins_\xi A_{(2)}  +{\rm c.c.}   \right],
\end{eqnarray}
and
\begin{eqnarray}\label{IIB:ThetaLag}
- \ins_\xi \mathcal{L}\big |_{on-shell} &=& \frac{1}{9!} \left[ \frac{1}{\kappa_{10}^2} \varepsilon_{a_1\cdots a_9 a}\left( 
\frac{1}{48} g^a_{\phantom{a}b}(G_{(3)})_{cde}(\bar{G}_{(3)})^{cde} \right)\xi^b \right]\dd x^{a_1}\wedge \cdots \wedge \dd x^{a_9}\nonumber\\
&& + \frac{i}{2\kappa_{10}^2} G_{(3)}\wedge\bar{G}_{(3)}\wedge \ins_\xi C_{(4)}
+ \frac{i}{2\kappa_{10}^2}  \Big(C_{(4)}\wedge\bar{F}_{(3)} \wedge \ins_\xi F_{(3)} + {\rm c.c.} \Big) \,.
\end{eqnarray}
Notice that in \eqref{IIB:ThetaG}-\eqref{IIB:ThetaLag} we have opted to display some of the contributions explicitly in terms of the 9-form components. This is because most of these contributions add-on to build the equation of motion for the graviton \eqref{eq:einstein} and thus will not contribute to the final current (as displayed, it is easy to track these contributions). The only such contribution that survives after using the graviton equation of motion is the first term in \eqref{IIB:ThetaG} that can be rewritten as:
 \begin{eqnarray}\label{IIB:stardxi}
\frac{1}{9!}\left[ \frac{1}{\kappa_{10}^2} \varepsilon_{a_1\cdots a_9 a} \nabla_b \nabla^{[b}\xi^{a]} \right]\dd x^{a_1}\wedge \cdots \wedge \dd x^{a_9} =\frac{1}{2\kappa_{10}^2}\dd\star\dd\xi\,.
\end{eqnarray}

Adding all contributions \eqref{IIB:ThetaG}-\eqref{IIB:ThetaLag}, using the equations of motion \eqref{eq:einstein} and \eqref{eq:f5}, and \eqref{IIB:stardxi} one finally finds that the sympletic Noether current 9-form \eqref{IIB:current} is given by 
\begin{eqnarray}\label{IIB:currentFinal1}
J=\dd \widetilde{Q}_\xi\,,
\end{eqnarray}
where we have defined the Noether 8-form charge  
\begin{eqnarray}\label{IIB:currentFinal2}
\widetilde{Q}_\xi &\equiv& \frac{1}{2\kappa_{10}^2}\star \dd \xi +\frac{4}{\kappa_{10}^2} \star F_{(5)} \wedge \ins_\xi C_{(4)}  \nonumber\\
&& + \frac{1}{4\kappa_{10}^2}\Big[ \left( f\left( \star\, \bar{G}_{(3)}-\bar{B}\star G_{(3)}\right) - i \star F_{(5)} \wedge\bar{A}_{(2)} +2i \,C_{(4)}\wedge\bar{F}_{(3)} \right) \wedge \ins_\xi A_{(2)} + {\rm c.c.}  \Big] \,.
\end{eqnarray}
Consistent with the discussion below \eqref{IIB:current}, we find that on-shell the current $J$ is indeed conserved, $\dd J=\dd^2\widetilde{Q}_\xi=0$. 

So far, we have taken $\xi$ to be just a diffeomorphism generator. Until otherwise stated, we now take  $\xi$ to be a Killing vector field  \cite{wald1993black, iyer1994some, iyer1995comparison, wald2000general} or an asymptotically Killing vector field  \cite{anderson1996asymptotic, barnich2002covariant, barnich2003boundary, barnich2008surface, Compere:2007vx, Compere:2007az, Chow:2013gba}. 
Moreover, consider a solution of IIB supergravity. It depends on one or more parameters $m_k$ (say, with $k=1,\cdots$) and we are interested in considering variations $\delta$ along this moduli space of solutions.
The variation $\omega_\xi$ (in the moduli space) of the charge associated to $\xi$ is then  \cite{wald1993black, iyer1994some, iyer1995comparison, wald2000general, barnich2002covariant, barnich2003boundary, barnich2008surface, Compere:2007vx, Compere:2007az}
\begin{equation}\label{VarCharge0}
\omega_\xi \equiv \sum_i  \left[ \delta \widetilde{Q}_\xi (\phi_i) - \widetilde{Q}_{\delta \xi}(\phi_i) -\ins_\xi \theta(\phi_i,\delta \phi_i) -  E_\mathcal{L} \left( \mathcal{L}_\xi \phi_i, \delta \phi_i \right) \right].
\end{equation}
The last contribution vanishes when $\xi$ is a Killing vector, i.e. when $\mathcal{L}_\xi \phi_i =0$. We are interested inly in such cases so we set $E_\mathcal{L} \left( \mathcal{L}_\xi \phi_i, \delta \phi_i \right)=0$ in the rest of our discussion. 
The third contribution is the interior product of the sum of the boundary terms $\theta(\phi_i,\delta \phi_i)$ given in \eqref{IIB:bdryG}, \eqref{IIB:bdryB}, \eqref{IIB:bdryA2} and \eqref{IIB:bdryA4}. 
Finally, the first and second contributions in \eqref{VarCharge0} are given by 
\begin{eqnarray}\label{VarCharge1}
&& \delta \widetilde{Q}_\xi =  \partial_{m_k} \!\widetilde{Q}_\xi \,  \dd m_k \,,\nonumber \\
&& \widetilde{Q}_{\delta \xi} = \widetilde{Q}_{\partial_{m_k}\!\xi \,  \dd m_k}\,.
\end{eqnarray}
Thus, $\widetilde{Q}_{\delta \xi}$ vanishes when the Killing vector $\xi^a$ is independent of the solution parameters $m_k$. This is certainly the case for the Killing vector fields $\xi=\partial_t$ and $\xi=\partial_{\psi_j}$ or $\xi=\partial_{w_j}$ responsible for time, rotational or translational symmetries, respectively. Therefore, $ \widetilde{Q}_{\delta \xi}=0$ for the cases we are interested. 

To sum, for each Killing vector field $\xi$ describing time, translational or rotational isometries, one can associate a 8-form Noether charge whose variation, in the moduli space of solutions, is given by 
 \begin{eqnarray}\label{IIB:VarCharge}
 \omega_\xi&=& \sum_k \left( \partial_{m_k} \!\widetilde{Q}_\xi \, \dd m_k \right)   \\
 && -\ins_\xi \theta(g,\delta g) -\ins_\xi \theta\left( B,\bar{B},\delta B,\delta\bar{B}\right)-\ins_\xi\theta\left(A_{(2)},\bar{A}_{(2)},\delta A_{(2)},\delta\bar{A}_{(2)}\right)-\ins_\xi \theta\left( C_{(4)}, \delta C_{(4)} \right), \nonumber
 \end{eqnarray}
with $\widetilde{Q}_\xi $ given in \eqref{IIB:currentFinal2} and the several boundary terms $\theta(\phi_i,\delta \phi_i)$  given by \eqref{IIB:bdryG}, \eqref{IIB:bdryB}, \eqref{IIB:bdryA2} and \eqref{IIB:bdryA4}, respectively. As a check of our computation, one can explicitly confirm, using the equations of motion (\ref{eqs:IIB}) and the fact that $\xi$ is a Killing vector, that $\omega_{\xi}$ is indeed a closed $8-$form, that is to say $\mathrm{d}\omega_{\xi}=0$. 

Equation \eqref{IIB:VarCharge} is an universal expression within IIB supergravity and one of our main contributions. It gives the variation (in the moduli or parameter space) of the conserved Noether charge associated to a Killing vector field $\xi$ (when $\xi$ is independent of the moduli). Note, however, that we have not yet discussed the physical interpretation for the associated Noether charge. For example, we might be tempted but we cannot say that $ \omega_\xi$ with $\xi=\partial_t$ describes the variation of the energy; in general it does not. To find this physical interpretation and the energy of a solution we need some physical input from the system at hand. We illustrate the usefulness of \eqref{IIB:VarCharge} for two distinct systems in the two next subsections. In the first case the system is manifestly integrable, meaning that the integration of \eqref{IIB:VarCharge} (with $\xi=\partial_t$) yields immediately the energy. In the second case, an inspection of \eqref{IIB:VarCharge} will suggest that we cannot define an energy for the system. However, using the asymptotic scale-invariance of the system we will be able to compute the energy and thermodynamic potential of the system unambiguously.

\subsection{Lumpy and Localized AdS$_5\times $S$^5$ Black Holes \label{sec:lumpyLoc}}
In this subsection we use \eqref{IIB:VarCharge} to find the energy of solutions of type IIB supergravity that are asymptotically AdS$_5\times S^5$ and that were previously constructed in \cite{dias2015lumpy, dias2016localized}. They describe lumpy and localized black holes, i.e. black holes that are deformed along the polar direction of the $S^5$. The equations of motion allow such a solution if the AdS$_5$ and $S^5$ radii, which we will denote by $L$, are the same. In these solutions, the only fields that are present are the graviton $g$ and the RR self-dual 5-form $F_{(5)}=\mathrm{d} C_{(4)}$, i.e. $B=0$ and $A_{(2)}=0$.    

A general ansatz for such a solution which is static, preserves the $SO(4)$ symmetry of AdS$_5$ and the $SO(5)$ subgroup of the S$^5$ (while eventually breaking its $SO(6)$ symmetry) and that has horizon topology $S^8$ is  \cite{dias2015lumpy}
\begin{subequations}
\label{IIB:lumpyansatz}
\begin{align}
& \d s^2 = \frac{L^2}{\left(1-y^2\right)^2}\Bigg[-y^2 \left(2-y^2\right)\,G(y)\,Q_1 \,\d t^2  \nonumber\\
& \hspace{3cm} +\frac{4 y_+^2}{\left(2-y^2\right) G(y)}\,Q_2 \,\left[\d y + (1-y^2)^2 Q_3 \,\d x\right]^2+y_+^2 Q_5 \, \d\Omega_3^2 \Bigg] \nonumber\\
& \hspace{1cm}  +L^2 \left[Q_4\,\frac{4 \d x^2}{2-x^2}+ Q_6\, \left(1-x^2\right)^2 \d\Omega_4^2 \right] \,,\\
& C_{(4)}=\frac{L^4 y_+^4}{\sqrt{2}} \frac{y^2 \left(2-y^2\right)}{\left(1-y^2\right)^4}\, H(y) \,Q_7 \, \d t\wedge \d S_{(3)}-\frac{L^4}{\sqrt{2}}\, W\,\d S_{(4)} \,.
\end{align}
\end{subequations}
where $\left\{Q_I,W\right\}$ ($I=1,\ldots, 7$) are functions of the $S^5$ polar coordinate $x$ and of the radial coordinate $y$ and, in addition, of the (single) black hole parameter $y_+$ that essentially gives the temperature of the black hole. 
One has $x\in[-1,1]$, with $x=\pm 1$ corresponding to the north and south poles of the S$^5$, and $y\in[0,1]$ with $y=0$ being the horizon location and $y=1$ being the boundary of the AdS$_5$.

 If we set $Q_1=Q_2=Q_4=Q_5=Q_6=Q_7=W=1$ and $Q_3=0$, we recover the familiar global AdS$_5$-Schwarzschild$\times$S$^5$ solution which preserves the $SO(6)$ symmetry group of the $S^5$. This is best seen if we apply the coordinate transformations $r=\frac{r_+ }{1-y^2}\,,\tilde{x}=x\sqrt{2-x^2}$  (and $y+=r_+/L$) which allow to rewrite the solution in its standard form, namely
\begin{subequations}
\label{IIB:Schw}
\begin{equation}
 \mathrm{d}s^2 = -f \mathrm{d}t^2+\frac{\mathrm{d}r^2}{f} + r^2 \mathrm{d}\Omega_3^2 +L^2 \left( \frac{\mathrm d \tilde x^2}{1-\tilde x^2}+(1-\tilde x^2)\d\Omega_4^2 \right) \,,
\end{equation}
\begin{equation}
F_{\mu \nu \rho\sigma\tau} = \epsilon_{\mu \nu \rho\sigma\tau}, \qquad
F_{abcde} = \epsilon_{abcde}\,,
\end{equation}
\begin{equation}
\hbox{where} \quad f=1+\frac{r^2}{L^2}-\frac{r_+^2}{r^2}\left(\frac{r_+^2}{L^2}+1\right) \,, \nonumber
\end{equation}
\end{subequations}
and $\epsilon_{\mu \nu \rho\sigma\tau} \, \mathrm{d}y^\mu\wedge \cdots \wedge \mathrm{d}y^\tau$ and $\epsilon_{abcde} \, \mathrm{d}x^a \wedge\cdots  \wedge \mathrm{d}x^e$ are the volume forms of the AdS$_5$ and $S^5$ base spaces, respectively.
 
Besides including the AdS$_5$-Schwarzschild$\times$S$^5$ as a special solution, the ansatz \eqref{IIB:lumpyansatz} also describes black hole solutions  that preserve the full $SO(4)$ symmetry of the S$^3$ but only an $SO(5)$ symmetry of the S$^5$, that is to say, that allow deformations along the polar direction $x$ that break $SO(6)$ down to $SO(5)$.  
For such `{\it lumpy}' black hole solutions one has $\left\{Q_I,W\right\}=\left\{Q_I(x,y;y_+),W(x,y;y_+)\right\}$ but it is still true that the AdS$_5$ or S$^5$ radius $L$ drops out of the equations of motion, so that the black hole is  described by the single parameter $y_+$. The temperature of the black hole is \cite{dias2015lumpy} 
\begin{equation}
T=\frac{1}{L}\frac{2 y_+^2+1}{2 \pi  y_+} \,.
\end{equation}

This background is a particularly good example to illustrate the practical use of \eqref{IIB:VarCharge} to obtain straightforwardly the energy of a `simple' system that depends on a single parameter ($y_+$) and for which the variation of the Noether charge is manifestly integrable. For $\xi=\partial_t$, \eqref{IIB:VarCharge} reduces in the present case to
\begin{eqnarray}\label{IIB:VarChargeLumpy}
\omega_\xi &=& \partial_{y_+} \!\widetilde{Q}_\xi \, \dd y_+  -\ins_\xi \theta(g,\delta g) -\ins_\xi \theta\left( C_{(4)}, \delta C_{(4)} \right)\nonumber\\
&=& \frac{1}{2\kappa_{10}^2} \Big\{ \partial_{y_+}\Big[ \star \dd \xi +8 \star F_{(5)} \wedge \ins_\xi C_{(4)}  \Big]  \nonumber \\
&& - \ins_\xi \Big[  \left( \nabla^b  \partial_{y_+} g_{ab}-g^{bc}\nabla_a  \partial_{y_+} g_{bc}\right) dx^a - 8 \star F_{(5)}\wedge  \partial_{y_+} C_{(4)} \Big] \Big\} \, \dd y_+
 \end{eqnarray}
where we collected  $\theta(g,\delta g)$ and $\theta\left( C_{(4)}, \delta C_{(4)} \right)$ from \eqref{IIB:bdryG} and \eqref{IIB:bdryA4}, respectively. 

To proceed, we need to borrow the asymptotic expansion of the fields $\left\{Q_I(x,y;y_+),W(x,y;y_+)\right\}$ from (A.7)-(A.10) of Appendix A of  \cite{dias2015lumpy} and that we do not reproduce here\footnote{Note that the 5-form used in \cite{dias2015lumpy} has a different normalization to the one used here, namely:  $F_{(5)}\big|_{\hbox{\cite{dias2015lumpy}}} =2\sqrt{2} F_{(5)}$.}.
This boundary expansion (up to the order where further contributions no longer contribute to the energy) depends on three undetermined coefficients that we promote to be functions of $y_+$: $\{\beta_2(y_+),\delta_{0}(y_+),\delta_4(y_+) \}$\footnote{As explained in detail in \cite{dias2015lumpy}, this is true after imposing appropriate Dirichlet boundary conditions that correspond to having no sources in the dual CFT$_4$ and that eliminate other integration constants.}. 
Inserting this asymptotic expansion of the fields into \eqref{IIB:VarChargeLumpy}, and integrating over a spacelike surface $S_t^\infty$ at constant time and at the asymptotic radial boundary, $y=1$, one gets 
\begin{eqnarray}\label{IIB:VarChargeLumpy2}
 \int_{S_t^\infty} \omega_{\partial_t} &=& \frac{\pi^4 L^8 y_+}{6144 G_{10}} \Big[ -4608 \left(1+2 y_+^2\right) +
5 y_+^2 \Big(4 \beta_2 (y_+)+24 \beta_2 (y_+)^2+ y_+ [1+12 \beta_2 (y_+)] \beta_2 '(y_+)\Big)\nonumber\\
&& \:\qquad\qquad+
192 y_+^2 \left[y_+ \delta_0 '(y_+)+4 \delta_0 (y_+)\right]+12 y_+^2 \left[y_+ \delta_4'(y_+)+4 \delta_4 (y_+)\right)
\Big]\dd y_+\,.
 \end{eqnarray}
Finally, to get the energy we integrate this along $y_+$:
\begin{equation} \label{IIB:energyLumpy}
E= \int dy_+ \int_{S_t^\infty} \omega_{\partial_t}=\frac{\pi^4 L^8}{6144 G_{10}}{\biggl [}2304 \,y_+^2(1+y_+^2) -y_+^4 {\biggl (}5\, \beta_2+30 \,\beta_2^2+12\left(16 \,\delta_{0}+\delta_4\right){\biggr)}{\biggr ]} +C_0.
\end{equation}
where $C_0$ is an integration constant to be fixed. In particular, if we set $\beta_2=\delta_0=\delta_4=0$ we get the energy of the global AdS$_5$-Schwarzschild$\times$ S$^5$ black hole,  $E=\frac{3 \pi^4L^8}{8G_{10}}\,y_+^2 \left(y_+^2+1\right)+C_0$.  
The constant $C_0$ is the energy assigned to the  global AdS$_5\times S^5$ background ($y_+=0$). The temperature and entropy of the lumpy black hole solutions can be read directly from \eqref{IIB:lumpyansatz}. The horizon quantities that appear in the entropy as well as  the asymptotic  parameters $\{\beta_2,\delta_{0},\delta_4 \}$ present in the energy \eqref{IIB:energyLumpy} can be 
extracted from the numerical solution of \cite{dias2015lumpy} via a fit of the Taylor expansion of the fields to the numerical data. Doing so one finds that the first law $\dd E=T \dd S$ is obeyed, as it should.

Equation \eqref{IIB:energyLumpy} further agrees with the expression for the energy one gets (see Appendix A of \cite{dias2015lumpy})  using the formalism of Kaluza-Klein holography and holographic renormalisation \cite{skenderis2006kaluza} (see also \cite{kim1985mass, gunaydin1985spectrum, lee1998three, lee1999ads5, arutyunov2000some, skenderis2006holographic, skenderis2007anatomy})\footnote{Note that Kaluza-Klein holography has the added value that it also gives the expectation values of other observables like the vevs of Kaluza-Klein scalars that condensate and that have a dual holographic interpretation \cite{skenderis2006kaluza, dias2015lumpy}}. Finally, we note that expression \eqref{IIB:energyLumpy}  gives also the correct energy for the {\it localized} AdS$_5\times S^5$ black hole of \cite{dias2016localized}. These are black holes that localize in the pole of the S$^5$, breaking totally the $SO(6)$ but not the $SO(5)$ symmetry of the S$^5$.  

\subsection{Polchinski-Strassler Black Brane \label{sec:PSbh}}
In this subsection we shall find the energy of the Polchinski-Strassler (PS) black hole constructed in \cite{Bena:2018vtu}. This is an asymptotically AdS$_5\times S^5$ solution of IIB supergravity where all the bosonic fields ($g,B,\bar{B},A_{(2)},\bar{A}_{(2)},C_{(4)}$) are non-trivial. The solution depends on two parameters, namely the dimensionless horizon radius $y_+$ and a supersymmetric mass deformation parameter $m$. The former  parameter essentially fixes the temperature of the black hole as 
\begin{equation}
T = \frac{y_+}{\pi L} \,,
\end{equation} 
with $L$ being the $AdS_5$ radius. The mass deformation parameter $m$ controls the magnitude of the asymptotic fall-off of the complex 3-form $G_{(3)}$. In short, deforming the theory with this source (i.e. boundary condition) is of interest because it introduces confinement in the holographic dual theory. The dual (3+1)-dimensional ${\cal N}=4$ SYM becomes confined if one deforms it by giving masses to the fermions and scalars, which are directly related to the supergravity parameter $m$. Typically these mass deformations break all the supersymmetries of the original SYM. However, there are certain configurations that preserve supersymmetry, yielding a dual deformed SYM theory which is known as the ${\cal N}=1^{*}$ theory. The PS black hole constructed in \cite{Bena:2018vtu} corresponds to the high-temperature phase of the confined Polchinski-Strassler vacuum \cite{polchinski2000string} of the ${\cal N}=1^{*}$ theory. 

This brief description of the PS black hole immediately suggests that it is a two-parameter solution, which depends on the horizon radius $y_+$ and the mass deformation parameter $m$ (see the discussion below for a revision of this statement). Having numerically constructed this solution in the earlier paper \cite{Bena:2018vtu}, we would like to be able to evaluate the energy and the chemical potential which is conjugate to the mass deformation $m$ from the asymptotic of the supergravity fields.\footnote{In particular, note that the procedure of holographic renormalization  to compute the thermodynamic quantities and expectation values of gravitational solutions with dual gauge theories is highly non-trivial for this system and is not available. Indeed, we might be tempted  to use the formalism of Kaluza-Klein holography and holographic renormalisation to find the thermodynamic observables \cite{kim1985mass, gunaydin1985spectrum, lee1998three, lee1999ads5, arutyunov2000some, skenderis2006kaluza, skenderis2006holographic, skenderis2007anatomy, dias2015lumpy}. However, so far Kaluza-Klein holography was developed only up to second order in the perturbative expansion of the fields and determining the thermodynamics of the PS black hole would require extending this analysis to fourth order. This is an absolutely non-trivial task.} Unfortunately, in this case it is not obvious that \eqref{IIB:VarCharge} is integrable and it is certainly not clear how we can use it to compute the energy and mass deformation potential from it. The additional ingredient needed to make the variation integrable, as we will show next, comes from holography.

\subsubsection{Thermodynamic Implications of Asymptotic Scale Invariance}
The Noether charge formalism applies to any solution of the supergravity theory. In the present case, the solution has the additional property of being holographic, so that it provides a dual gravitational description of some field theory. The black brane solution is asymptotically scale-invariant, which corresponds to the statement that the dual field theory is conformal in the UV. The energy and other conserved charges in conformal field theories transform covariantly under the dilatation operator, and as a consequence of holography, we can conclude that the  conserved charges in the dual gravitational description must transform in the same way. By directly imposing these transformation properties, and manipulating relations that they imply, we will be able to show that for the Polchinski-Strassler black brane the Noether variation does indeed become integrable. 

To make this discussion concrete, consider planar AdS$_5$ in Fefferman-Graham coordinates $\{z,x^\mu\}$, with $x^\mu=\{t, w^i\}$. Here $t$ is the time coordinate, $w^i$ with $j \in \{1,2,3\}$ are the translationally invariant planar directions of AdS$_5$, and $z$ is the holographic radial coordinate (with the conformal boundary being at $z=0$). In these coordinates, planar AdS$_5$ is manifestly invariant under dilatations of the Lorentz coordinates, $x^{\mu} \to \lambda \,x^{\mu}$ if we also scale the radial coordinate as $z\to \lambda\,z$. This scale-invariance implies that any  thermodynamic quantity must be a homogeneous function of the scale factor $\lambda$.\footnote{Recall that a function $f$ is homogeneous if $f(\lambda x) = \lambda^k f(x)$ for arbitrary real $\lambda$.} For example, the temperature $T$ is essentially the inverse of the Euclidean time circle length and thus has mass dimension 1: under a dilatation one has $T\to T/\lambda$. This is also the mass dimension of the mass deformation $m$ since asymptotically $G_{(3)}$ decays as $m \,z$ \cite{Bena:2018vtu}. The entropy density $s$ (i.e. the entropy divided by the volume of the three planar directions $w^i$) has mass dimension 3. Lastly, by the first law, the energy density $\rho$ must have the same mass dimension as the product of the temperature by the entropy, namely 4. \footnote{Alternatively,  $\rho$ has the same mass dimension has the time-time component of the holographic stress tensor which, in a 4-dimensional holographic boundary, is 4.}.

From this simple analysis we can extract a few far-reaching conclusions. First, although the solution ansatz depends on two parameters, $T$ and $m$, any gauge invariant physical property of the solution must be scale-invariant and therefore only depends on the ratio $\hat{m} \equiv m/T$. Equivalently, if we parametrize the solution by $y_+$ and $m$, we can use scale-invariance to fix $y_+=1$, then a change in $m$ moves us along the parameter space.  Second, from a physical perspective, the mass deformation parameter $m$ is fixed asymptotically, and the energy density is then computed at fixed $m$ (and $y_+$). Hence, in the microcanonical ensemble, the energy density $\rho$ should  be a function only of the entropy density $s$ and $m$: $\rho = \rho(s,m)$. The homogeneous properties of these thermodynamic quantities imply that the  scale invariant quantities are $m/T, s/T^3$ and $\rho/T^4$. It follows  that under a scale transformation the energy density must respect
\begin{equation}
\rho(\lambda^{3} s,\lambda m)=\lambda^4\rho(s,m)\,.
\label{IIB:scaling}
\end{equation}

Another important conclusion follows from the fact that $\rho$, $s$, and $m$ are the extensive variables of the system. Thus, the variations of these extensive quantities  should be related by  a first law of thermodynamics of the form
\begin{equation}
\delta \rho = T\,\delta s+\vartheta \,\delta m\,,
\label{IIB:firstlaw}
\end{equation}
where, the temperature $T$ is the conjugate intensive variable to $s$ and we have defined $\vartheta$ to be the conjugate intensive variable to $m$.\footnote{Here, $\delta$ represents a variation along the moduli space of solutions. The moduli space is spanned by $(y_+, m)$, and $\delta$ is taken to be a linear variation in an arbitrary direction, i.e. $\delta = \delta y_+ \partial_{y_+} + \delta m\,\partial_{m}$, with $\delta y_+, \delta m$ arbitrary. Note that we have not yet used scale-invariance to set $y_+ = 1$ at this stage, which would cut down the dimension of the moduli space to 1.}. Finally, the two relations \eqref{IIB:scaling} and \eqref{IIB:firstlaw} allow to deduce a Smarr relation for our PS black holes as follows. We start by taking a derivative of  \eqref{IIB:scaling} with respect to $\lambda$, and then set $\lambda = 1$ to get
$$
3 s \,\frac{\partial \rho}{\partial s}+m \,\frac{\partial \rho}{\partial m}=4 \,\rho(s,m)\,.
$$
We now use the first law \eqref{IIB:firstlaw} to compute the partial derivatives in terms of $T$ and $\vartheta$, and this yields the desired Smarr relation:
\begin{equation}
T\,s = \frac{4}{3}\rho-\frac{1}{3}\vartheta\,m\,.
\label{IIB:smarr}
\end{equation}

To summarize, the holographic interpretation and the scale invariant properties of our system implies that the energy density $\rho$ of our solutions must satisfy the first law  \eqref{IIB:firstlaw}  and the Smarr relation  \eqref{IIB:smarr}. This is valuable new information which will allow to integrate the Noether charge variation \eqref{IIB:VarCharge}, and thus compute the energy density $\rho$ and mass deformation potential $\vartheta$ of the PS black hole of  \cite{Bena:2018vtu}. We do this in the next subsection.

\subsubsection{Calculation of the Energy and Chemical Potential\label{sec:PSenergy}}
In order to carry out the calculation of the energy and chemical potential, we will need to have the Taylor expansion around the horizon and asymptotic boundary, of the fields of the PS black hole (subject to the appropriate physical boundary conditions). To avoid unnecessary repetition of information, we ask the reader to consult equations (4.34) of the companion paper \cite{Bena:2018vtu}. There, we give the bosonic fields $(g, B, \bar{B}, A_{(2)}, \bar{A}_{(2)}, C_{(4)})$ for the Polchinski-Strassler black brane that asymptotes to $AdS_5\times S^5$ and has horizon topology $\mathbb{R}^3\times S^5$. This is a cohomogeneity-2 solution depending on a polar coordinate $x$ along the $S^5$ and on a compact radial coordinate $y \in [0,1]$. The horizon is located at $y=0$ and the conformal boundary is at $y=1$. The solution is described by a total of 20 independent functions, labelled as $q_j(x,y)$ ($j=1,2,\cdots,20$) in \cite{Bena:2018vtu} and that obey  boundary conditions (vanishing sources for all the operators at the boundary and regularity at the horizon) also detailed in \cite{Bena:2018vtu}, namely in its equation (4.36) and below.
To be clear, the original fields were redefined in terms of $q_j(x,y)$ as a way to impose more straightforwardly the boundary conditions and to easy the numerical computation.  

The field \emph{ansatz} (4.34) of \cite{Bena:2018vtu} is written in coordinates which explicitly manifest the Killing symmetries of the system, namely time translation with Killing vector field $\xi=\partial_t$, and spatial translation with $\xi_i = \partial_{w_i}$ ($i\in\{1,2,3\}$) along the brane directions. Hence, we can construct constant $t$ and $w_i$ hypersurfaces which we identify as $\Sigma_{t}$ and $\Sigma_{w_i}$, respectively, or collectively as $\Sigma_\xi$. These are 9-dimensional hypersurfaces over which we can integrate the exterior derivative of the 8-form $\omega_\xi$  which, recall, is closed, $\dd \omega_\xi =0$. For integration along  hypersurfaces of constant $t$ we take the coordinates $w_i$ to be periodic with period $\Delta w_i$, and for the hypersurfaces of constant $w_i$ we take $t\in[0,\Delta t]$. In each of these 9-dimensional hypersurfaces $\Sigma_\xi$ we define a constant $y$ slice that we call $S^y_{\Sigma_{\xi}}$. The boundary of $\Sigma_\xi$ is then given by the slice at $y=0$ and the slice at $y=1$. Integrating $\mathrm{d}\omega_{\xi}$, given by \eqref{IIB:VarChargeLumpy}, over these hypersurfaces yields 
\begin{equation}
\label{IIB:smarrdiff}
0=\int_{\Sigma_{\xi}} \mathrm{d} \omega_{\xi} = \int_{S^{y=1}_{\xi}} \omega_{\xi}-\int_{S^{y=0}_{\xi}} \omega_{\xi}\qquad \Rightarrow \qquad \int_{S^{y=1}_{\xi}} \omega_{\xi}=\int_{S^{y=0}_{\xi}}\omega_{\xi}\,,
\end{equation}
where we used Stoke's theorem to get the last equality. 

Borrowing the Taylor expansion of the fields at the horizon ($y=0$) from \cite{Bena:2018vtu}, we evaluate the integral on the right-hand side of \eqref{IIB:smarrdiff} for $\xi=\partial_t$ and find that it is proportional to  $T \delta s$,
\begin{equation}
\label{eq:bigone}
\frac{1}{\Delta w_1\Delta w_2\Delta w_3}\int_{S^{y=0}_{\partial_t}}\omega_{\partial_t}= T \delta s \,.
\end{equation}
From \eqref{IIB:smarrdiff}-\eqref{eq:bigone} and from the first law \eqref{IIB:firstlaw}, it follows that the left-hand side of \eqref{IIB:smarrdiff}, which is evaluated at the conformal boundary, must be proportional to
\begin{eqnarray}\label{IIB:bigone}
 \delta{\cal I}_{\rm 1st  \, law}^{\infty} &=& T \delta s\,,  \qquad \qquad \hbox{with} \qquad  \delta{\cal I}_{\rm 1st  \, law}^{\infty}\equiv  \frac{1}{\Delta w_1\Delta w_2\Delta w_3} \int_{S^{y=1}_{\partial_t}}\omega_{\partial_t}\,, \nonumber \\
&=& \delta\rho-\vartheta \,\delta m\,.
\end{eqnarray}
This asymptotic integral $ \delta{\cal I}_{\rm 1st \, law}^{\infty}$ is a {\it known} function of the asymptotic decays (including non-leading terms) of some of the fields $(g, B, \bar{B}, A_{(2)}, \bar{A}_{(2)}, C_{(4)})$ $-$ that we borrow from \cite{Bena:2018vtu} $-$ which {\it are themselves a function of the phase space parameters $(y_+,m)$}. That is to say, $\delta{\cal I}_{\rm 1st  \, law}^{\infty}=\delta{\cal I}_{\rm 1st  \, law}^{\infty} (y_+,m)$. Similarly, $\rho=\rho(y_+,m)$ and $\vartheta=\vartheta(y_+,m)$.
Equation \eqref{IIB:bigone}, with roots in the first law, will prove to be one of the two fundamental relations that we will use below.

Proceeding and again borrowing the Taylor expansion of the fields around the horizon from \cite{Bena:2018vtu}, we observe that the following difference of integrals evaluated at the horizon hypersurface holds
\begin{equation}
\frac{1}{\Delta w_1\Delta w_2\Delta w_3}\int_{S^{y=0}_{\partial_t}}\omega_{\partial_t}-\frac{1}{\Delta t}\int_{S^{y=0}_{\partial_{w_i}}}\omega_{\partial_{w_i}} = \delta (T s)\,.
\end{equation}
It then follows from \eqref{IIB:smarrdiff} that the same integral difference, when evaluated at the asymptotic hypersurface $(y=1)$, must give the same result, i.e.
\begin{eqnarray}
\frac{1}{\Delta w_1\Delta w_2\Delta w_3}\int_{S^{y=1}_{\partial_t}}\omega_{\partial_t}-\frac{1}{\Delta t}\int_{S^{y=1}_{\partial_{w_i}}}\omega_{\partial_{w_i}} &=& \frac{1}{\Delta w_1\Delta w_2\Delta w_3}\int_{S^{y=0}_{\partial_t}}\omega_{\partial_t}-\frac{1}{\Delta t}\int_{S^{y=0}_{\partial_{w_i}}}\omega_{\partial_{w_i}} \nonumber\\
&=& \delta (T s)\,.
\end{eqnarray}
This implies $-$ via the variation of the Smarr relation \eqref{IIB:smarr}, $\delta (T s)= \delta \left(\frac{4}{3}\,\rho-\frac{1}{3}\vartheta\,m\right)$ $-$ that 
\begin{equation} \label{IIB:bigtwo}
\delta{\cal I}_{\rm Smarr}^{\infty}  =\delta \left(\frac{4}{3}\,\rho-\frac{1}{3}\vartheta\,m\right), 
\qquad \hbox{with} \quad   \delta{\cal I}_{\rm Smarr}^{\infty} \equiv  \frac{1}{\Delta w_1\Delta w_2\Delta w_3}\int_{S^{y=1}_{\partial_t}}\omega_{\partial_t}-\frac{1}{\Delta t}\int_{S^{y=1}_{\partial_{w_i}}}\omega_{\partial_{w_i}} \,.
\end{equation}
This asymptotic integral $ \delta{\cal I}_{\rm Smarr}^{\infty}=\delta{\cal I}_{\rm Smarr}^{\infty}(y_+,m)$ is also {\it known} function of the asymptotic decays (including non-leading terms) of some of the fields $(g, B, \bar{B}, A_{(2)}, \bar{A}_{(2)}, C_{(4)})$ $-$ that we borrow from \cite{Bena:2018vtu} $-$ which {\it are themselves a function of the phase space parameters $(y_+,m)$}. 
Equation \eqref{IIB:bigtwo}, that traces back its origin to the Smarr relation, the second fundamental relation we shall use below.

Armed with these first law and Smarr relations \eqref{IIB:bigone} and \eqref{IIB:bigtwo} (which emerge from the asymptotic scale invariance of our system), we may now calculate the energy $\rho(y_+,m)$ and mass deformation conjugate potential $\vartheta(y_+,m)$ of the PS black brane from the knowledge of $\delta{\cal I}_{\rm 1st  \, law}^{\infty}(y_+,m)$ and $\delta{\cal I}_{\rm Smarr}^{\infty}(y_+,m)$. Again, these two integrals can be explicitly computed from the asymptotic decay of the fields $(g, B, \bar{B}, A_{(2)}, \bar{A}_{(2)}, C_{(4)})$ (i.e. $q_j(y)$ with $j \in \{1,\dotsc,20\}$) determined in \cite{Bena:2018vtu}. 

To accomplish our task, we have to consider a general variation in the moduli space spanned by $(y_+, m)$.
To begin, equation  \eqref{IIB:bigone} $-$ $\delta{\cal I}_{\rm 1st  \, law}^{\infty}  = \delta \rho-\vartheta \, \delta m$ $-$ when viewed as a function of $(y_+, m)$ can be rewritten as
\begin{eqnarray}\label{IIB:bigone2}
\big(\partial_{y_+}   \delta{\cal I}_{\rm 1st  \, law}^{\infty} \big) \dd y_+ +  \big(\partial_{m}   \delta{\cal I}_{\rm 1st  \, law}^{\infty} \big)  \dd m=\big(\partial_{y_+}\rho \, \dd y_+  +\partial_{m}\rho \, \dd m\big)  -\vartheta \, \dd m\,.
\end{eqnarray}
The $\dd y_+$ and $\dd m$ contributions to this variation must vanish independently which gives the following relations for $\partial_{y_+}\rho$ and $\partial_{m}\rho$: 
\begin{equation}\label{IIB:bigoneFinal}
 \left\{
\begin{array}{ll}
 \partial_{y_+}\rho = \partial_{y_+}\delta{\cal I}_{\rm 1st  \, law}^{\infty} , &  \\
 \partial_{m}\rho =  \partial_{m}\delta{\cal I}_{\rm 1st  \, law}^{\infty}   + \vartheta\,.&
\end{array}
\right. 
\end{equation}

Now take again  \eqref{IIB:bigtwo}, $\delta{\cal I}_{\rm Smarr}^{\infty}  =\delta \left(\frac{4}{3}\,\rho-\frac{1}{3}\vartheta\,m\right)$. Considering its variation along the  moduli space parametrized by $(y_+, m)$ one gets
\begin{equation}\label{IIB:bigtwoFinal}
\big(\partial_{y_+}\delta{\cal I}_{\rm Smarr}^{\infty}\big) \dd y_+ +  \big(\partial_{m} \delta{\cal I}_{\rm Smarr}^{\infty}\big) \dd m 
= \frac{4}{3} \big(\partial_{y_+}\rho \, \dd y_+  +\partial_{m}\rho \, \dd m\big)  -\frac{1}{3}\big(\partial_{y_+}\vartheta \, \dd y_+  +\partial_{m}\vartheta \, \dd m\big). 
\end{equation}
As before, the $\dd y_+$ and $\dd m$ contributions of this variation must vanish independently. Moreover, we can insert \eqref{IIB:bigoneFinal} into \eqref{IIB:bigtwoFinal}. Altogether, we end up with a system of two coupled non-homogeneous first order equations for $\vartheta(y_+,m)$,
\begin{equation}\label{IIB:FinalPDEs}
 \left\{
\begin{array}{ll}
 m\, \partial_{y_+}\vartheta=  4 \partial_{y_+}\delta{\cal I}_{\rm 1st  \, law}^{\infty} -3 \partial_{y_+}\delta{\cal I}_{\rm Smarr}^{\infty}, &  \\
m \partial_{m}\,\vartheta -3 \vartheta= 4 \partial_{m}\delta{\cal I}_{\rm 1st  \, law}^{\infty} -3 \partial_{m}\delta{\cal I}^{\infty}_{\rm Smarr}\,.&
\end{array}
\right. 
\end{equation}
which are sourced by $\delta{\cal I}_{\rm 1st  \, law}^{\infty}(y_+,m)$ and $\delta{\cal I}^{\infty}_{\rm Smarr}(y_+,m)$ which are known functions given by the asymptotic integrals \eqref{IIB:bigone} and \eqref{IIB:bigtwo}, respectively. These expressions are not illuminating so we do not display them. We note, however, that solutions $\vartheta(y_+,m)$ of Eq.~(\ref{IIB:FinalPDEs}) only exist if $\delta{\cal I}_{\rm 1st  \, law}^{\infty} $ and $\delta{\cal I}_{\rm Smarr}^{\infty}$ are such that $\partial_{\mu}\partial_{y_+}\vartheta$ read from the first equation in (\ref{IIB:FinalPDEs}) yields the same result as $\partial_{y_+}\partial_{\mu}\vartheta$ read from the second equation in (\ref{IIB:FinalPDEs}). This \emph{integrability condition} is in essence what fails in \cite{Chow:2013gba}.

We can now solve \eqref{IIB:FinalPDEs} to find the mass deformation potential $\vartheta(y_+,m)$ and then insert this solution into the decoupled non-homogeneous first order equations \eqref{IIB:bigoneFinal} to find the energy density $\rho(y_+,m)$.
The final result, after writing the gravitational constant in terms of field theory quantities, $\kappa^2 \equiv 8 \pi G_{10}= 4 \pi ^5 L^8/N^2$, is \cite{Bena:2018vtu}:
\begin{eqnarray} \label{IIB:energy}
&& \hspace{-0.5cm} \vartheta = -\frac{3 N^2}{2 \pi ^2} \left[y_+^2\,\beta _1(m,y_+)+C_1 m^3+\frac{14 m^2}{9}\right],\\
&&  \hspace{-0.5cm} \rho = -\frac{N^2}{48 \pi ^2} \Big[6 \,y_+^4\alpha _0(m,y_+)+18 C_1 m^4+3 m^4+28 m^3+3 m^2 y_+^2+18 \beta _1(m,y_+) m\,y_+^2-18\,y_+^4+\rho_0\Big]. \nonumber
\end{eqnarray}
Here, $\rho_0$ and $C_1$ are the two integration constants of the problem and they will be determined below using supersymmetry considerations. On the other hand, $\alpha_0(y_+,m)$ and $\beta_1(y_+,m)$ are expansion parameters of the fields around the asymptotic boundary which we now discuss. 

For that first recall that in the gauge used in \cite{Bena:2018vtu}, the PS black brane of is described by a total of 20 independent functions $q_j(x,y)$ ($j=1,2,\cdots,20$) identified in the field \emph{ansatz}  (4.34) of \cite{Bena:2018vtu}. For our purposes, we are interested in solving the equations of motion in an asymptotic expansion around the conformal boundary located at $y=1$. To do so we check that all our 20 fields have an asymptotic expansion in powers of $(1-y)$ (up to the order we need to read the thermodynamics of the system): 
\begin{equation}\label{IIB:expansion}
q_j(x,y)=\sum_{J=0}^{2} a_i^{(J)}(x)(1-y)^J+o[(1-y)^{2}]\,, \qquad (i=1,2,\cdots,20).
\end{equation}
Note that there are no logarithmic terms in this expansion up to order $(1-y)^2$ in agreement with the expectations discussed in \cite{taylor2001anomalies} for this system. The energy (and mass deformation potential) only depends on contributions of \eqref{IIB:expansion} up to order $(1-y)^2$. This can be anticipated using dimension counting. Indeed, the stress energy tensor has conformal dimension $4$, and asymptotically one has $1-y\propto z^2$ (where $z$ is the Fefferman-Graham coordinate). Therefore, the terms that contribute to the energy are those that appear to order $(1-y)^2\propto z^4$.

To describe the origin of the parameters $\alpha_0$ and $\beta_1$ that appear in \eqref{IIB:energy}, we just need to look into the asymptotic expansion of three of these functions, namely $q_1,q_{11}$ and $q_{20}$. Consulting the field \emph{ansatz} (4.34) of \cite{Bena:2018vtu}, we see that $q_1$ is related to the time-time component $g_{tt}$ of the metric, $q_{11}$ is one of the 8 (independent) components of $A_{(2)}$ and $q_{20}$ is one of the two (independent) components of $C_{(4)}$. Up to the relevant order, the asymptotic Taylor expansion of these three fields is
\begin{subequations}\label{IIB:expansion3qs}
\begin{align}
&q_1(x,y)=1+\frac{m^2}{y_+^2}(1-y)+\left[\alpha_0(m,y_+)-\frac{4\,m^4}{y_+^4} \left(\frac{1-x^2}{1+x^2}\right)^2\right](1-y)^2+o[(1-y)^2]\,,
\\
&q_{11}(x,y)=m+\left[\beta_1(m,y_+)+\frac{m^2 \left(7-24 m+7 x^2\right)}{3\,y_+^2 \left(1+x^2\right)}\right](1-y)+o[(1-y)]\,,
\\
&q_{20}(x,y)=1-\frac{2\,m^2}{y_+^2} (1-y)+\alpha _2(m,y_+)(1-y)^2+o[(1-y)^2]\,,
\end{align}
\end{subequations}
which explicitly contains the parameters $\alpha_0(m,y_+)$, $\beta_1(m,y_+)$ and $\alpha_2(m,y_+)$ that are constants (dependent on $m$ and $y_+$). These constants can be extracted from the numerical functions via a fit of \eqref{IIB:expansion3qs} to the numerical data. This concludes the interpretation of the parameters  $\alpha_0$ and $\beta_1$ in \eqref{IIB:energy}.

Next, consider the integration constants $\rho_0$ and $C_1$. These constants are both fixed by supersymmetry. The $\mathcal{N} = 1^*$ theory at hand has one supersymmetric solution that is explicitly known, namely the 10-dimensional uplift \cite{Petrini:2018pjk,Bobev:2018eer},\cite{Bena:2018vtu} of the 5-dimensional ${\cal N}=8$ gauge supergravity solution of Giradello, Petrini, Porrati and Zaffaroni (GPPZ) \cite{Girardello:1999bd} (the fact that this solution is singular is irrelevant for our purposes). Naturally, we require that such a supersymmetric solution has zero energy density (this agrees with the holographic renormalization computation of the GPPZ energy density done in \cite{Bianchi:2001kw,Bianchi:2001de},\cite{Bena:2018vtu}). This condition requires that we choose $\rho_0=0$ and $C_1 = -4/3$ in \eqref{IIB:energy}.

This concludes the calculation of the entropy and chemical potential of the PS black brane. As there are many steps involved, we will summarize the steps that lead to the unambiguous definition of energy density and mass deformation conjugate potential \eqref{IIB:energy}. Using the canonical energy formalism of Wald an expression for the variation for the conserved Noether charge \eqref{IIB:VarCharge} was derived. This expression was not integrable in general, and {\it \`a priori} it was unclear if it could be used to define an unambiguous energy density for the PS black brane solution. We then showed that asymptotic scale-invariance allows two additional relations to be derived, a first law \eqref{IIB:firstlaw} and a Smarr relation \eqref{IIB:smarr}. These two thermodynamic relations, when combined with the Noether charge variation \eqref{IIB:VarCharge}, lead to the fundamental relations \eqref{IIB:bigone} and \eqref{IIB:bigtwo}. These relations then lead to a system of first order PDEs \eqref{IIB:bigoneFinal} and \eqref{IIB:FinalPDEs} for $\vartheta(y_+, m)$ and $\rho(y_+, m)$. Finally, these PDEs may be integrated to yield \eqref{IIB:energy}, with $\rho_0=0$ and $C_1 = -4/3$, which give the energy density $\rho$ and mass deformation conjugate potential $\vartheta$ as a function of the known asymptotic decay of the fields that describe the PS black brane.

\section{11-Dimensional Supergravity \label{sec:Mtheory}}
In this section, we will repeat the analysis of the previous section but this time for 11-dimensional supergravity.

The bosonic fields of  11-dimensional supergravity are the metric field $g$ and a 3-form gauge potential $A_{(3)}$ with associated field strength  $G_{(4)}=\mathrm{d}A_{(3)}$. The associated action is
\begin{equation}
S_{11} = \frac{1}{2 \kappa_{11}^2} \int \bigg( \star \mathcal{R} - \frac{1}{2} \, G_{(4)} \wedge \star\, G_{(4)} + \frac16 \, G_{(4)} \wedge G_{(4)} \wedge A_{(3)} \bigg)\,,
\label{M:action}
\end{equation}
where $\mathcal{R}$ is the Ricci volume form, $\mathcal{R}=R\,\star \mathbb{I}=R\,\mathrm{vol}_{11}$. 

\subsection{Variation of the Noether Charge}
As in the IIB case, we are interested in taking the variation of action \eqref{M:action} to get the equations of motion as well as the associated boundary terms that will be fundamental to compute the charges of solutions of the theory. The variation of \eqref{M:action} with respect to the graviton yields:
\begin{eqnarray}\label{M:varG}
 \delta S_{11}\big |_g &=& \frac{1}{2 \kappa_{11}^2} \int \mathrm{d}^{11}x\sqrt{-g}\left[ R_{ab}-\frac{1}{2} \,R g_{ab}-\frac{1}{12}\left({G_{(4)}}_{acde} {G_{(4)}}_b{}^{cde} - \frac{1}{8} g_{ab} {G_{(4)}}_{cdef} {G_{(4)}}^{cdef}\right)\right] \delta g^{ab} \nonumber\\
 && + \delta S_{g}^{\text{GH})}
\end{eqnarray}
with the Gibbons-Hawking boundary term being 
\begin{equation}\label{M:bdryG0}
 \delta S_{g}^{\text{GH})}=\frac{1}{2 \kappa_{11}^2} \int \mathrm{d}^{11}x \sqrt{-g} \,g^{ab}\,\delta R_{ab} =\frac{1}{2 \kappa_{11}^2} \int \mathrm{d}^{11}x \sqrt{-g} \, \nabla^a  \left( \nabla^b \delta g_{ab}-\nabla_a \delta g\right).  
\end{equation}
From this, the associated Gibbons-Hawking 10-form may be straightforwardly found to be
\begin{equation}\label{M:bdryG}
\theta(g,\delta g)=\frac{1}{2 \kappa_{11}^2}  \star\left[ \left( \nabla^b \delta g_{ab}-\nabla_a \delta g\right) \dd x^a \right].  
\end{equation}
Note that the last term in \eqref{M:action} is a Chern-Simmons term. Hence, it does not depend on $\sqrt{-g}$ and does not contribute to the equation of motion for the metric $g$. Lastly, the variation with respect to the gauge field $A_{(3)}$ gives:
\begin{eqnarray}\label{M:varA3}
 \delta S_{11}\big |_{A_{(3)}} &=&  \frac{1}{2 \kappa_{11}^2} \int \delta \big |_{A_{(3)}}\left[ -\frac{1}{2} G_{(4)}\wedge \star\,G_{(4)} +\,\frac{1}{6}\, G_{(4)}\wedge G_{(4)} \wedge A_{(3)}\right] \nonumber \\
 &=& \frac{1}{2 \kappa_{11}^2} \int\left( \mathrm{d} \star G_{(4)}-\frac{1}{2}G_{(4)}\wedge G_{(4)}  \right) \wedge \delta A_{(3)}
 +   \theta\left( A_{(3)}, \delta A_{(3)} \right) \,,
\end{eqnarray}
with the boundary term given by 
\begin{equation}\label{M:bdryA3}
 \theta\left( A_{(3)}, \delta A_{(3)} \right)=\frac{1}{2\kappa_{11}^2} \left(\star \,G_{(4)}-\frac{1}{3}\, G_{(4)}\wedge A_{(3)} \right)\wedge \delta A_{(3)}\,.
\end{equation}

Requiring that the variations \eqref{M:varG} and \eqref{M:varA3} vanish yields the equations of motion of 11-dimensional supergravity:
\begin{subequations}
\label{eqs:M}
\begin{align}
&R_{ab} = \frac{1}{12}\left[{G_{(4)}}_{acde} {G_{(4)}}_b{}^{cde} - \frac{1}{12} g_{ab} \,{G_{(4)}}_{cdef} {G_{(4)}}^{cdef}\right]\,, \label{M:einstein}
\\
&\mathrm{d} \star  G_{(4)} = \frac{1}{2} G_{(4)} \wedge G_{(4)},
\label{M:maxwell}
\end{align}
\end{subequations}
where we contracted the equation of motion that follows from \eqref{M:varG} with the inverse metric to get the Ricci scalar $R$ and then inserted this quantity back into  \eqref{M:varG} to get the trace reversed equation of motion for the graviton \eqref{M:einstein}.

It follows from \eqref{M:einstein} that the on-shell Ricci volume form is $\star\mathcal{R}=\frac{1}{3!}G_{(4)}\wedge\star G_{(4)}$ which we insert into \eqref{M:action}  to get the on-shell 11-form Lagrangian,
\begin{equation}\label{M:onshellL}
\mathcal{L}\big |_{on-shell} = \frac{1}{2 \kappa_{11}^2} \left( -\frac{1}{3} G_{(4)}\wedge \star \,G_{(4)} +\frac{1}{6}\, G_{(4)} \wedge G_{(4)} \wedge A_{(3)} \right).
\end{equation}

The sympletic Noether current 10-form associated to a diffeomorphism vector generator $\xi$ is then \cite{wald1993black, iyer1994some, wald2000general}
 \begin{eqnarray}\label{M:current}
J=\Theta\left(g,\mathcal{L}_\xi g\right)+\Theta\left(A_{(3)}, \,\mathcal{L}_\xi A_{(3)}\right)- \ins_\xi \mathcal{L}\big |_{on-shell} \,,
\end{eqnarray}
where, as before, $\Theta(\phi_i,\mathcal{L}_\xi \phi_i)\equiv \theta(\phi_i,\mathcal{L}_\xi \phi_i)$ but this time for $\phi_i = \{g,A_{(3)} \}$. Recall that $\mathcal{L}_\xi \phi_i$ is the  Lie derivative of the field $\phi$ along $\xi$, and $ \ins_\xi \mathcal{L}\big |_{on-shell} $ is the interior product  of $\xi$ with the 11-form \eqref{M:onshellL}. 

Further recall that, given the equations of motion $E_i$ written in \eqref{eqs:M}, it can be shown that $\dd J=- E_i \,\mathcal{L}_\xi \phi_i$ (summation convention is assumed) \cite{wald1993black, iyer1994some, wald2000general}. Therefore the on-shell ($E_i=0$) current is closed, $\dd J=0$, for all $\xi$. It follows that there is a Noether charge 9-form $\widetilde{Q}_\xi$, locally constructed from $\{\xi, \phi_i\}$, such that on-shell one has $J=\dd \widetilde{Q}_\xi$ \cite{wald1993black,iyer1994some,wald2000general}.

To evaluate each term in \eqref{M:current} we use the identities listed in \eqref{Cartan} and in \eqref{pformIdentities} as well as the equations of motion \eqref{eqs:M}. One finds:
 \begin{eqnarray}\label{M:ThetaG}
 \Theta\left(g,\mathcal{L}_\xi g\right)&=&\star\Big( \frac{1}{\kappa_{11}^2}\left[\nabla^b \nabla_{(a}\xi_{b)} -g^{bc}\nabla_a\nabla_{(b}\nabla_{c)}\right]\dd x^a \Big)\nonumber\\
 &=&\frac{1}{10!}\left[ \frac{1}{\kappa_{11}^2} \varepsilon_{a_1\cdots a_{10} a}\left( \nabla_b \nabla^{[b}\xi^{a]}+R^a_{\:\:b}\xi^b\right)\right]\dd x^{a_1}\wedge \cdots \wedge \dd x^{a_{10}}\,, 
\end{eqnarray}
 \begin{eqnarray}\label{M:ThetaA4}
  \Theta\left( A_{(3)}, \mathcal{L}_\xi A_{(3)} \right)&=& \frac{1}{2\kappa_{11}^2} \left(\star \,G_{(4)}-\frac{1}{3}\, G_{(4)}\wedge A_{(3)} \right) \wedge \left[ \ins_\xi \dd A_{(3)} + \dd \left(\ins_\xi A_{(3)} \right) \right] \nonumber\\
&=& \frac{1}{10!} \left[ \frac{1}{\kappa_{11}^2} \varepsilon_{a_1\cdots a_{10} a}\left( 
-\frac{1}{12} (G_{(4)})^{a c_1c_2 c_3}(G_{(4)})_{b c_1c_2 c_3}\right)\xi^b \right]\dd x^{a_1}\wedge \cdots \wedge \dd x^{a_{10}}\nonumber\\
&& -\frac{1}{6\kappa_{11}^2} \Big[G_{(4)}\wedge G_{(4)}\wedge\ins_\xi A_{(3)}+ G_{(4)}\wedge A_{(3)} \wedge\ins_\xi  G_{(4)} \Big]\nonumber\\
&& - \frac{1}{\kappa_{11}^2}\, \dd\left[ \left( \star \,G_{(4)} -\frac{1}{3}\,G_{(4)} \wedge  A_{(3)} \right) \wedge \ins_\xi A_{(3)} \right]+\frac{1}{2\kappa_{11}^2} \dd \star G_{(4)}  \wedge \ins_\xi A_{(3)} \,, 
\end{eqnarray}
and 
 \begin{eqnarray}\label{M:ThetaLag}
- \ins_\xi \mathcal{L}\big |_{on-shell} &=& \frac{1}{10!} \left[ \frac{1}{\kappa_{11}^2} \varepsilon_{a_1\cdots a_{10} a}\left( 
\frac{1}{2}\frac{1}{3} \frac{1}{4!} g^a_{\phantom{a}b}(G_{(4)})_{c_1c_2c_3c_4}(G_{(4)})^{c_1c_2c_3c_4} \right)\xi^b \right]\dd x^{a_1}\wedge \cdots \wedge \dd x^{a_{10}}\nonumber\\
&& + \frac{1}{2\kappa_{11}^2} \left[ \frac{1}{3}\,G_{(4)}\wedge A_{(3)}\wedge \ins_\xi G_{(4)}
-\frac{1}{6} \,G_{(4)}\wedge G_{(4)} \wedge \ins_\xi A_{(3)} \right].
\end{eqnarray}
 In \eqref{M:ThetaG}-\eqref{M:ThetaLag} we display some of the contributions explicitly in terms of the 10-form components because most of these contributions add-on to build the equation of motion for the graviton \eqref{M:einstein} and thus will not contribute to the final current. The only exception is the first term in \eqref{M:ThetaG} which can be rewritten as:
 \begin{eqnarray}\label{M:stardxi}
\frac{1}{10!}\left[ \frac{1}{\kappa_{11}^2} \varepsilon_{a_1\cdots a_{10} a} \nabla_b \nabla^{[b}\xi^{a]} \right]\dd x^{a_1}\wedge \cdots \wedge \dd x^{a_{10}} =\frac{1}{2\kappa_{11}^2}\dd\star\dd\xi\,.
\end{eqnarray}

 Adding the three contributions \eqref{M:ThetaG}-\eqref{M:ThetaLag} and using the equations of motion \eqref{eqs:M} one finally finds that the (on-shell conserved) sympletic Noether current 10-form \eqref{M:current} is given by
 \begin{eqnarray}\label{M:currentFinal1}
J=\dd \widetilde{Q}_\xi\,,
\end{eqnarray}
where we have defined the Noether 9-form charge  
 \begin{equation}\label{M:currentFinal2}
\widetilde{Q}_\xi \equiv \frac{1}{2\kappa_{11}^2} \left[ \star\, \dd \xi -\left( \star \, G_{(4)}-\frac{1}{3} G_{(4)}\wedge A_{(3)}\right)\wedge \ins_\xi A_{(3)}  \right].
\end{equation}

We now let the diffeomorphism generator $\xi$  be a Killing vector field  \cite{wald1993black, iyer1994some, iyer1995comparison, wald2000general} or an asymptotically Killing vector field  \cite{anderson1996asymptotic, barnich2002covariant, barnich2003boundary, barnich2008surface, Compere:2007vx, Compere:2007az, Chow:2013gba}. A generic solution of 11-dimensional supergravity depends on one or more parameters $m_k$ (say, $k=1,\cdots$) and we want to consider variations $\delta$ along this moduli space of solutions. The associated variation $\omega_\xi$ (in the moduli space) of the charge associated to $\xi$ is given by \eqref{VarCharge0}, with $ \delta \widetilde{Q}_\xi $  and $\widetilde{Q}_{\delta \xi} $ defined in \eqref{VarCharge1} \cite{wald1993black, iyer1994some, iyer1995comparison, wald2000general, barnich2002covariant, barnich2003boundary, barnich2008surface, Compere:2007vx, Compere:2007az}.
As in the IIB case, onwards we restrict our attention to cases where: 1) $\xi$ is a Killing vector (which implies $\mathcal{L}_\xi \phi_i =0$) and 2) this Killing vector $\xi$ is independent of the solution parameters $m_k$ (which implies that $\widetilde{Q}_{\delta \xi}=0$).

It follows that, for each Killing vector field $\xi$, one can associate a 9-form Noether charge whose variation in the moduli space of solutions is given by 
 \begin{equation}\label{M:VarCharge}
 \omega_\xi= \sum_k \left( \partial_{m_k} \!\widetilde{Q}_\xi \, \dd m_k \right) -\ins_\xi \theta(g,\delta g) -\ins_\xi \theta\left( A_{(3)}, \delta A_{(3)} \right),
 \end{equation}
with $\widetilde{Q}_\xi $ given in \eqref{M:currentFinal2} and the boundary terms $\theta(\phi_i,\delta \phi_i)$  given by \eqref{M:bdryG} and \eqref{M:bdryA3}. Using the equations of motion (\ref{eqs:M}) and the fact that $\xi$ is a Killing vector, we can check that $\omega_{\xi}$ is indeed a closed $9-$form, $\mathrm{d}\omega_{\xi}=0$. 

Equation \eqref{M:VarCharge} is a universal expression within 11-dimensional supergravity and is one of our main contributions. It gives the variation (in the moduli space) of the conserved Noether charge associated to a Killing vector field $\xi$. In the following, we will illustrate how this expression can be used to calculate conserved charges for specific solutions. We will first consider the  CGLP black brane recently constructed in the companion paper \cite{Dias:2017opt}. 

\subsection{CGLP Black Brane\label{subsec:CGLPbh}}
The  CGLP black brane recently constructed in the companion paper \cite{Dias:2017opt} is a solution of \eqref{M:action} which is asymptotically $AdS_4\times V_{5,2}$, where $V_{5,2}$ is the 7-dimensional Stiefel manifold. It is the simplest finite temperature  solution that asymptotes to the Cveti\v{c}, Gibbons, L\"u, and Pope (CGLP) solution, which has a magnetic $G_{(4)}$-flux background whose asymptotic decay describes a mass deformation of the corresponding dual $CFT_3$ \footnote{The CGLP \cite{cvetivc2003ricci} solution is a smooth, supersymmetric (and thus zero-temperature) solution to 11-dimensional supergravity describing ``fractional'' M2-branes dissolved in flux, which generalizes the Klebanov-Strassler solution \cite{klebanov2000supergravity} to higher dimensions. The CGLP is to be seen as the confined phase of the (poorly understood) dual $CFT_3$ theory while the CGLP brane of \cite{Dias:2017opt} is to be interpreted as the deconfined dual phase.}. At a linear level, this solution can be constructed starting with the asymptotically $AdS_4\times V_{5,2}$ Schwarzschild black brane (which has a purely electric $G_{(4)}$, i.e. with components only along the $AdS_4$ directions) and adding a linearized mass deformation (i.e., a magnetic contribution to $A_{(3)}$ which has components along the $V_{5,2}$) \cite{Dias:2017opt}\footnote{The Schwarzschild black brane describes the near-horizon geometry of finite-temperature M2-branes at the singular tip of the Stiefel cone.}. 

As suggested in the above description, the CGLP black brane of \cite{Dias:2017opt} depends on two parameters: the dimensionless horizon radius $y_+$, which fixes the temperature of the solution as 
\begin{equation}
T = \frac{3 y_+}{4\pi L} \,,
\end{equation} 
with $L$ being the $AdS_4$ radius, and a mass deformation parameter $\mu$ which can be read from the leading order decay of the magnetic part of the $G_{(4)}$-flux  \cite{Dias:2017opt}. This a cohomogeneity-1 solution: it only depends on a radial coordinate that in \cite{Dias:2017opt} is taken to be a compact coordinate $y\in [0,1]$, with $y=0$ being the location of the conformal boundary and $y=1$ being the location of the Killing horizon.\footnote{\label{foot}Note that the location of the horizon and conformal boundary of the Polchinsky-Strassler solutions of section \ref{sec:PSenergy} \cite{Bena:2018vtu} were {\it instead} at $y=0$ and $y=1$, respectively. That is, the physical roles of $y=0$ and $y=1$ were traded w.r.t. to the CGLP solutions of this section \ref{subsec:CGLPbh}  \cite{Dias:2017opt}.}  An {\it ansatz} for the metric $g$ and gauge $A_{(3)}$ fields of this CGLP black brane is given in equations (5.3) of \cite{Dias:2017opt}. It depends on 9 functions $q_j(y)$, with $j \in \{1,\dotsc,9\}$ which are determined in \cite{Dias:2017opt} solving the equations of motion \eqref{eqs:M} in the so-called DeTurck gauge and subject to the physically relevant boundary conditions also detailed in \cite{Dias:2017opt}.

Our aim in this section is to derive an expression for the energy of the CGLP black brane from the supergravity fields $q_j(y)$ found in \cite{Dias:2017opt}. However, the main obstacle encountered in the Polchinski-Strassler black brane analysis of section \ref{sec:PSbh} is also encountered here: the variation of the Noether charge \eqref{M:VarCharge} is not integrable as it is. Not surprisingly, the same technique that worked earlier will also resolve this problem. Indeed, the CGLP black brane is asymptotically $AdS_4\times V_{5,2}$ and thus conformal to $M^{1,3}$. Consequently, the dual field theory preserves conformal invariance in the UV and, in particular, it is thus invariant under scale (dilatation) transformations. The asymptotic scale-invariance of the gravitational solution, combined with the implication of holography that physical quantities may be read off from the boundary, will lead to additional relations (first law and Smarr) which, when imposed, render the Noether charge variation integrable. 

To fully exploit scale-invariance consider first the mass dimensions of the thermodynamics quantities which must be homogeneous functions of the scaling factor. Under a dilatation with scaling factor $\lambda$, the temperature $T$ (i.e. the inverse of the Euclidean time circle length) has mass dimension 1, i.e. $T\to T/\lambda$. The entropy density $s$ (i.e. the entropy divided by the volume of the two planar directions $w^i$ of $AdS_4$) has mass dimension 2, $s\to s/\lambda^2$. Asymptotically, the magnetic contribution of the potential $A_{(3)}$ decays as $A_{(3)} \sim \mu \,z^{2/3}$, where $z$ is the  Fefferman-Graham radial coordinate and $\mu$ is the mass deformation. Therefore the mass deformation has mass dimension $2/3$ under a dilatation. By the first law, the energy density $\rho$ must have mass dimension 3, i.e. the same dimension as the product of the temperature by the entropy (of course, this is also the mass dimension of the time-time component of the holographic stress tensor which, in a 3-dimensional holographic boundary, is 3). 

The asymptotic scale-invariance of our solution has four far-reaching consequences.
First we can use it to fix $y_+=1$ and then move in parameter space by changing $\mu$, since only the ratio $\hat{\mu}\equiv \mu/T^{2/3}$ is scale invariant. Second, from a physical viewpoint, the mass deformation $\mu$ is a parameter that we fix asymptotically, and the energy density is then computed at fixed $\mu$ (and $y_+$). That is to say, in the microcanonical ensemble, the energy density $\rho$ should  be a function only of the entropy density $s$ and $\mu$: $\rho = \rho(s,\mu)$. The homogeneous properties of these thermodynamic quantities then implies that under a scale transformation one has
\begin{equation}
\rho(\lambda^2 s,\lambda^{2/3}\mu)=\lambda^3\rho(s,\mu)\,.
\label{M:scaling}
\end{equation}
A third consequence of scale-invariance is that much like the entropy density $s$ is the thermodynamic conjugate variable to the temperature $T$, the mass deformation $\mu$ must also have a thermodynamic conjugate variable $-$ a density that we shall call the mass deformation potential $\vartheta$.  So, the energy density is  a homogeneous function of only $s$ and $\mu$ and these have  conjugate variables $T$ and $\vartheta$, respectively. Hence, we should have a first law of the form
\begin{equation}
\delta \rho = T \, \delta s+\vartheta\,\delta \mu\,,
\label{M:first}
\end{equation}
where $\delta$ denotes a variation along the moduli space of solutions, in our case, $\delta = \delta y_+\partial_{y_+} + \delta \mu\partial_{\mu}$.\footnote{Technically, in order to use the consequences of scale-invariance and the conserved Noether charge to find the energy, it is important that at this stage we do not use scale-invariance to fix $y_+=1$, as will become clear in the discussions after \eqref{M:bigone} and  \eqref{M:bigtwo}.} The fourth and final consequence is that the scaling relation \eqref{M:scaling} and the first law \eqref{M:first} imply a Smarr relation. Indeed, taking a derivative of \eqref{M:scaling} with respect to $\lambda$, and then setting $\lambda = 1$ yields the Smarr relation:
\begin{equation}\label{M:Smarr}
2\,s\,\frac{\partial \rho}{\partial s}+\frac{2}{3}\mu\frac{\partial \rho}{\partial \mu}=3\rho(s,\mu) \quad \Rightarrow \quad  T s = \frac{3}{2}\rho-\frac{1}{3}\vartheta\,\mu\,,
\end{equation}
where the last implication follows from the first law \eqref{M:first}.

Using the first law \eqref{M:first} and the Smarr relation \eqref{M:Smarr}, the variation of the Noether charge \eqref{M:VarCharge} may be shown to be integrable. First, note that the field {\it ansatz} for the CGLP black brane $-$ see equations (5.3) of \cite{Dias:2017opt} $-$ is written in coordinates that explicitly indicate that $\xi=\partial_t$ and $\xi=\partial_{w_i}$ ($i=1,2$) are Killing vector fields (these 3 directions,  parametrize the world-volume directions of the CGLP brane). We denote hypersurfaces of constant $t$ by $\Sigma_t$ and those of constant $w_i$  by $\Sigma_{w_i}$. These are $10-$dimensional hypersurfaces. Therefore we can integrate $\mathrm{d}\omega_{\partial_t}$ and $\mathrm{d}\omega_{\partial_{w_i}}$ over $\Sigma_t$ and $\Sigma_{w_i}$, respectively. For the  integration over $\Sigma_{w_i}$ we take  $t\in[0,\Delta t]$, while for the integration over $\Sigma_t$ we take the coordinates $w_i$ to be periodic with period $\Delta w_i$. In either case, we can define  $S^y_{\Sigma_{\xi}}$ to be a constant $y$ slice of $\Sigma_{\xi}$ and use of Stoke's theorem yields,
\begin{equation}
0=\int_{\Sigma_{\xi}} \mathrm{d} \omega_{\xi} = \int_{S^{y=0}_{\xi}} \omega_{\xi}-\int_{S^{y=1}_{\xi}} \omega_{\xi}\quad \Rightarrow \quad \int_{S^{y=0}_{\xi}} \omega_{\xi}=\int_{S^{y=1}_{\xi}}\omega_{\xi}\,.
\label{M:smarrdiff}
\end{equation}
The right hand side of \eqref{M:smarrdiff}, i.e. at the horizon hypersurface $S^{y=1}_{\partial_t}$, for $\xi=\partial_t$ can be evaluated inserting the Taylor expansion around the horizon of the fields $\{g(y),A_{(3)}(y)\}$ found in \cite{Dias:2017opt} into \eqref{M:VarCharge}. We find that it is proportional to $T \delta s$, 
\begin{equation}\label{M:equalw}
\frac{1}{\Delta w_1\Delta w_2}\int_{S^{y=1}_{\partial_t}}\omega_{\partial_t} = T \delta s\,.
\end{equation}
It then follows from \eqref{M:smarrdiff}-\eqref{M:equalw} and from the first law \eqref{M:first} that the left hand side of \eqref{M:smarrdiff}, which is evaluated at the asymptotic hypersurface $S^{y=0}_{\partial_t}$, must be given by
\begin{eqnarray}\label{M:bigone}
{\cal I}_{\rm 1st\,law}^{\infty}&=& T \delta s\,, \qquad\qquad \hbox{with} \qquad {\cal I}_{1st\,law}^{\infty}\equiv \frac{1}{\Delta w_1\Delta w_2}\int_{S^{y=0}_{\partial_t}}\omega_{\partial_t}\,, \nonumber\\
&=& \delta \rho-\vartheta \, \delta \mu\,,
\end{eqnarray}
where the asymptotic integral $ \delta{\cal I}_{\rm 1st \, law}^{\infty}=\delta{\cal I}_{\rm 1st  \, law}^{\infty} (y_+,\mu)$ is a {\it known} function of the asymptotic decays (including non-leading terms) of some of the fields $\{g(y),A_{(3)}(y)\}$ $-$ that we borrow from \cite{Dias:2017opt} $-$ which {\it are themselves a function of the phase space parameters $(y_+,\mu)$}. Similarly, $\rho=\rho(y_+,\mu)$ and $\vartheta=\vartheta(y_+,\mu)$. This relation \eqref{M:bigone} is one of two key relations that we use below.

Next, inserting their Taylor expansion of the fields $\{g(y),A_{(3)}(y)\}$ found in \cite{Dias:2017opt} around the horizon into \eqref{M:VarCharge}, we find that the following  relation holds at the horizon hypersurface ($y=1$):
\begin{equation}
\frac{1}{\Delta w_1\Delta w_2}\int_{S^{y=1}_{\partial_t}}\omega_{\partial_t}-\frac{1}{\Delta t}\int_{S^{y=1}_{\partial_{w_i}}}\omega_{\partial_{w_i}} = \delta (T s)\,.
\end{equation}
But \eqref{M:smarrdiff}  requires that the same integral difference also holds when evaluated at the asymptotic hypersurface $S^{y=0}_\xi$:
\begin{eqnarray} \label{M:diffInt}
\frac{1}{\Delta w_1\Delta w_2}\int_{S^{y=0}_{\partial_t}}\omega_{\partial_t}-\frac{1}{\Delta t}\int_{S^{y=0}_{\partial_{w_i}}}\omega_{\partial_{w_i}} &=&\frac{1}{\Delta w_1\Delta w_2}\int_{S^{y=1}_{\partial_t}}\omega_{\partial_t}-\frac{1}{\Delta t}\int_{S^{y=1}_{\partial_{w_i}}}\omega_{\partial_{w_i}} \nonumber\\
&=& \delta (T s)\,,
\end{eqnarray}
which implies, via the Smarr relation \eqref{M:Smarr}, that
\begin{equation}
\delta{\cal I}_{\rm Smarr}^{\infty} =\delta \left(\frac{3}{2}\rho-\frac{1}{3}\vartheta\,\mu\right),\qquad \hbox{with} 
\qquad \delta{\cal I}_{\rm Smarr}^{\infty}  \equiv \frac{1}{\Delta w_1\Delta w_2}\int_{S^{y=0}_{\partial_t}}\omega_{\partial_t}-\frac{1}{\Delta t}\int_{S^{y=0}_{\partial_{w_i}}}\omega_{\partial_{w_i}}\,.
\label{M:bigtwo}
\end{equation}
Again, $\delta{\cal I}_{\rm Smarr}^{\infty}(y_+,\mu)$ is a {\it known} function of the asymptotic decays (including non-leading terms) of some of the fields $\{g(y),A_{(3)}(y)\}$ \cite{Dias:2017opt}  which are themselves a function of $(y_+,\mu)$. 
Equation \eqref{IIB:bigtwo} is a second fundamental relation we use below.

With the first law and Smarr relations \eqref{M:bigone} and \eqref{M:bigtwo} (originated from the asymptotic scale invariance of our system), we can now compute the energy density $\rho(y_+,\mu)$ and mass deformation conjugate potential $\vartheta(y_+,\mu)$ of the CGLP black brane from the asymptotic decay of the fields $\{g(y),A_{(3)}(y)\}$ (i.e. $q_j(y)$ with $j \in \{1,\dotsc,9\}$) determined in \cite{Dias:2017opt}. For this, we follow {\it mutatis mutandis} the procedure detailed in \eqref{IIB:bigone2}-\eqref{IIB:energy} but this time using the first law and Smarr relations \eqref{M:first}-\eqref{M:bigtwo} of the CGLP system (see footnote \ref{foot}).
Summarizing this procedure, we consider a general variation in the moduli space spanned by $(y_+, \mu)$. Equation \eqref{M:bigone} has two contributions, one for each of the variations $\dd y_+$ and $\dd \mu$, that must vanish independently. From these two conditions one  gets 
\begin{equation}\label{M:bigoneFinal}
 \left\{
\begin{array}{ll}
 \partial_{y_+}\rho = \partial_{y_+}\delta{\cal I}_{\rm 1st  \, law}^{\infty} , &  \\
 \partial_{\mu}\rho =  \partial_{\mu}\delta{\cal I}_{\rm 1st  \, law}^{\infty}   + \vartheta\,.&
\end{array}
\right. 
\end{equation}
that we replace into \eqref{M:bigtwo}. Again this yields two contributions (for $\dd y_+$ and $\dd \mu$) that must vanish independently. These give  two coupled non-homogeneous first order PDEs for $\vartheta(y_+,\mu)$:
\begin{equation}\label{M:FinalPDEs}
 \left\{
\begin{array}{ll}
 \mu\, \partial_{y_+}\vartheta=  \frac{9}{2}  \, \partial_{y_+}\delta{\cal I}_{\rm 1st  \, law}^{\infty} -3 \partial_{y_+}\delta{\cal I}_{\rm Smarr}^{\infty}, &  \\
\mu \,\partial_{\mu}\,\vartheta -\frac{7}{2}  \,  \vartheta= \frac{9}{2} \, \partial_{\mu}\delta{\cal I}_{\rm 1st  \, law}^{\infty} -3 \partial_{\mu}\delta{\cal I}^{\infty}_{\rm Smarr}\,.&
\end{array}
\right. 
\end{equation}
which are sourced by $\delta{\cal I}_{\rm 1st  \, law}^{\infty}(y_+,\mu)$ and $\delta{\cal I}^{\infty}_{\rm Smarr}(y_+,\mu)$ that are the known  asymptotic integrals \eqref{M:bigone} and \eqref{M:bigtwo}, respectively. 
We can now solve \eqref{M:FinalPDEs} to find the mass deformation potential $\vartheta(y_+,\mu)$ (see comments below Eq.~(\ref{IIB:FinalPDEs})) and then insert this solution into the decoupled non-homogeneous first order equations \eqref{M:bigoneFinal} to find the energy density $\rho(y_+,\mu)$.
Because their equations are sourced by $\delta{\cal I}_{\rm 1st  \, law}^{\infty}(y_+,\mu)$ and $\delta{\cal I}_{\rm Smarr}(y_+,\mu)$, $\vartheta(y_+,\mu)$ and $\rho(y_+,\mu)$ are a function of the asymptotic decay of the fields $q_1(y)$ and $q_7(y)$ determined in \cite{Dias:2017opt} ($q_1$ is proportional to the $g_{tt}$ component of the metric and $q_7$ is proportional to one of the components of the $3$-form $A_{(3)}$), namely:
\begin{subequations}\label{M:energyMu}
\begin{align}
&\vartheta = \frac{N^{3/2}}{4374 \pi ^6} \left[12 \pi  C_1 \mu ^{7/2}-5 y_+^{5/3} \hat{q}_7^{\prime\prime}(0)\right],
\\
&\rho = \frac{N^{3/2}}{419904 \pi ^6} \left[\rho_0+256 \pi  C_1 \mu ^{9/2}-192 \mu  y_+^{5/3} q_7''(0)-729 y_+^3 \left(3 q_1''(0)-4\right)\right].
\end{align}
\end{subequations}
The arbitrary integration constants $\rho_0, C_1$ can be fixed by applying these relations to the CGLP solution, and requiring that its energy density vanishes (because CGLP is a supersymmetric solution). This fixes $\rho_0 = 0$ and $C_1 = 0$ \footnote{For an alternative route to determine the integration constants, consider for a moment that the CGLP solution was not available. Even in these conditions, we could still argue that $C_1$ should vanish since it is natural to expect that the energy should be an analytic function of $\mu$. This argument is supported by the fact that we can set a perturbation theory around the Schwarzschild brane with $\mu=0$, whereby the magnetic $G_{(4)}$-flux fields only contain odd powers of $\mu$, and the remaining functions only have even powers of $\mu$: we ask the reader to see section 4 of \cite{Dias:2017opt} for details of this perturbative expansion.}.

It is worthwhile to summarize the path that led to the unambiguous definition of energy density and mass deformation conjugate potential \eqref{M:energyMu}. Using the canonical energy formalism of Wald and collaborators, we first derived the variation for the conserved Noether charge \eqref{M:VarCharge}. This expression was not integrable and it looked like that we could not define an energy density for our CGLP black brane. However, a key observation is that this particular solution is dual to a conformal field theory with a conformal UV fixed-point. As a result, the gravitational and field theory conserved charges agree, and both transform covariantly under the dilatation operator. Specifically, the energy and chemical potential obey a simple scaling law, which we used to derive a first law \eqref{M:first} and a Smarr relation for the system \eqref{M:Smarr}. These two thermodynamic relations then imply (via \eqref{M:bigone} and \eqref{M:bigtwo}) that the Noether charge variation is integrable, which allowed us to finally write the energy density $\rho$ and mass deformation conjugate potential $\vartheta$ in terms of the known asymptotic decay of the fields that describe the solution.

\section{Conclusions \label{sec:conc}}

Developing the notion, or the most appropriate and general notion, of mass and conserved charges in general relativity (including all its possible cosmological backgrounds) and associated formalisms to compute them has not been a straightforward path (as briefly reviewed in our introduction). Thus, it comes with no surprise that this task is even less trivial once we consider extensions of general relativity to include supergravity fields. 
Indeed, more often than not, finding the conserved thermodynamic quantities of supergravity solutions is a non-trivial task. In particular, when these are type IIB or eleven dimensional supergravities. Yet, these are the gauge invariant quantities of preference (not to say unique) to describe the solutions of the theory and to discuss preferred thermal phases of the phase diagram of solutions in a given thermodynamic ensemble. Sometimes, the particular solution at hand has a known dimensional reduction to a solution of a lower dimensional supergravity theory where one can use holographic renormalization, including Kaluza-Klein holographic renormalization, to compute  the desired thermodynamic quantities \cite{henningson1998holographic, balasubramanian1999stress, de2001holographic, bianchi2002holographic, bianchi2001go, de2000holographic, kalkkinen2001holographic, martelli2003holographic, papadimitriou2004ads, elvang2016practical, skenderis2006kaluza, kanitscheider2008precision, papadimitriou2005thermodynamics,elvang2016practical}. But this is definitely not always the case. Even if it is the case, Kaluza-Klein holographic renormalization is not (at least yet) fully developed to compute the thermodynamics of solutions where {\it all} type II supergravity or 11-dimensional SUGRA fields are turned on.

Fortunately, when one does not have the appropriate dimensional reduction and holographic renormalization tools, and we need to compute the thermodynamic quantities of a solution of the full blasted type II supergravity or 11-dimensional supergravity, one can resort to the covariant Noether charge formalism of Wald and collaborators \cite{wald1993black, iyer1994some, iyer1995comparison, wald2000general,papadimitriou2005thermodynamics} and its cohomological formalism extension \cite{Anderson:1996sc,Barnich:2001jy,Barnich:2003xg,Barnich:2007bf,Compere:2007vx,Chow:2013gba,Compere:2007az}.
In this context, two of our main results are given by \eqref{IIB:VarCharge}, for IIB SUGRA, and \eqref{M:VarCharge}, for 11D SUGRA. These equations give the universal expression  for the variation (in the moduli or parameter space) of the conserved Noether charge associated to a Killing vector field $\xi$ (when $\xi$ is independent of the moduli) in general IIB or 11D supergravity backgrounds, respectively. 

But even these covariant Noether charge and cohomological formalisms have a known limitation. Indeed, the covariant Noether charge formalism and its cohomological extension can certainly determine the  {\it variation} of a Noether charge along the moduli space of parameters that parametrize the phase space of a given family of solutions. When the solution at hand is a 1-parameter solution, we can certainly {\it integrate} this Noether variation to get the associated conserved Noether charge. This is {\it e.g.} illustrated in our IIB supergravity example of section \ref{sec:lumpyLoc} where one computes the energy of the asymptotically AdS$_5\times S^5$ lumpy  \cite{dias2015lumpy} and localized  \cite{dias2016localized} black holes of type IIB supergravity and finds that it matches the energy originally computed in \cite{dias2015lumpy,dias2016localized} using Kaluza-Klein holographic renormalization.

But when the solutions at hand are parametrized by 2 or more parameters, integrability of the  Noether charge variation cannot be taken as granted. There are many examples, where a system in these conditions is still integrable (most notably in the Kerr solution of general relativity). But there is also at least one know example \cite{Chow:2013gba} where it was declared that the Noether charge variation is not integrable and thus considered that the energy of this system is simply not defined.

In the present article, we observe that even when a multi-parameter solution is not manifestly integrable we can use {\it asymptotic scale invariance} to help us define certain asymptotic charges. Indeed, we have shown that when the system has {\it asymptotic scale invariance} we can still integrate the Noether charge variation to get the energy (density) of the system and other thermodynamic variables. This is because scale invariance implies that we can define a first law of thermodynamics and a Smarr relation. Altogether, these two extra conditions guarantee that we can integrate the Noether charge variation. We have explicitly demonstrated this is the case with two non-trivial examples of type IIB and 11-dimensional supergravity, namely:  the Polchinski-Strassler black brane of \cite{Bena:2018vtu} and the Cveti\v{c}-Gibbons-L\"u-Pope (CGLP) black brane of \cite{Dias:2017opt}.

The main results of our article are encapsulated in two rather compact expressions for the Noether charges of type IIB and 11-dimensional supergravities, which we can associate with solutions possessing exact Killing vectors $\xi$. These Killing vectors typically represent time, translational or rotational isometries of the solution in question. For type IIB we found
 \begin{multline}
 \label{IIB:VarChargea}
 \omega^{\mathrm{IIB}}_\xi=\sum_k \left( \partial_{m_k} \!\widetilde{Q}_\xi \, \dd m_k \right)   \\
-\ins_\xi \theta(g,\delta g) -\ins_\xi \theta\left( B,\bar{B},\delta B,\delta\bar{B}\right)-\ins_\xi\theta\left(A_{(2)},\bar{A}_{(2)},\delta A_{(2)},\delta\bar{A}_{(2)}\right)-\ins_\xi \theta\left( C_{(4)}, \delta C_{(4)} \right), \nonumber
 \end{multline}
 with $\widetilde{Q}_\xi $ given in \eqref{IIB:currentFinal2} and the several boundary terms $\theta(\phi_i,\delta \phi_i)$  given by Eqs.~(\ref{IIB:bdryG})-(\ref{IIB:bdryA4}), while for 11-dimensional supergravity we obtained
\begin{equation}
\label{M:VarChargea}
 \omega^{\mathrm{11D}}_\xi= \sum_k \left( \partial_{m_k} \!\widetilde{Q}_\xi \, \dd m_k \right) -\ins_\xi \theta(g,\delta g) -\ins_\xi \theta\left( A_{(3)}, \delta A_{(3)} \right)\,,\nonumber
 \end{equation}
with $\widetilde{Q}_\xi $ given in \eqref{M:currentFinal2} and the boundary terms $\theta(\phi_i,\delta \phi_i)$  given by \eqref{M:bdryG} and \eqref{M:bdryA3}. Knowledge of the 8-form $\omega^{\mathrm{IIB}}_\xi$ or 9-form $\omega^{\mathrm{11D}}_\xi$, together with asymptotic scale invariance, allowed us to determine the corresponding asymptotic charges in type IIB and 11-dimensional supergravity, respectively.

It is important to highlight that these three examples have in common the fact that they are solutions of IIB or 11D supergravity with all the possible supergravity fields switched on. Also important, they all have a holographic description, and asymptotic scale-invariance is a common property for gravitational solutions in the context of the gauge/gravity correspondence. Indeed, all  three spacetimes are asymptotically a direct product spacetime $AdS_p\times X^n$ where $X^n$ is a compact manifold. And this immediately invites the use of holographic renormalization techniques to compute the thermodynamic quantities. But in some of the cases, the holographic understanding that we have of the system is {\it not} yet sufficiently developed to be able to compute the energy and other thermodynamic quantities of the system. In particular, the powerful Kaluza-Klein holographic renormalization formalism is not yet developed to consider {\it any} system that has {\it all} supergravity fields non-vanishing. When this is the case, the use of the covariant Noether charge formalism combined with the  first law and Smarr relations $-$ that follow from the underlying asymptotic scale-invariance of the system $-$ is then a fundamental tool to compute the energy density and thermodynamic potentials. Of course, it would be interesting to use our systems as a testbed to develop the Kaluza-Klein holographic renormalization formalism further up to the point where it can also compute the thermodynamics of the solutions of \cite{Dias:2017opt,Costa:2014wya}. The direct holographic renormalization computation would have the added value of also identifying the expectation values of the several operators of the system (not only the conserved charges and thermodynamic potentials).  

\subsubsection*{Acknowledgements}
We wish to thank Benjamin E.~Niehoff for collaboration in the early stage of this project, and Iosif Bena and Benjamin E.~Niehoff for our collaborations in the partner projects \cite{Bena:2018vtu,Dias:2017opt}. We also thank Kostas Skenderis for useful discussions.  O.~J.~D. and G.~S.~H. acknowledge financial support from the STFC Ernest Rutherford grants ST/K005391/1 and ST/M004147/1 and from the STFC ``Particle Physics Grants Panel (PPGP) 2016" Grant No. ST/P000711/1. J.~E.~S. is supported in part by STFC grants PHY-1504541 and ST/P000681/1. J.~E.~S. also acknowledges support from a J. Robert Oppenheimer Visiting Professorship.

\appendix
\section{Useful Differential Form Relations}
\label{sec:diffForms}
For reference, here we include some useful definitions and identities for differential forms in a spacetime of dimension $d$ (and a single timelike direction). We will denote the Levi-Civita tensor as $\varepsilon_{a_1\cdots a_d}$, with orientation $\varepsilon_{123\cdots d}\equiv 1$. In the following, $P_{(p)}, A_{(p)}, B_{(p)}$ are $p$-forms, and $Q_{(q)}$ is a $q$-form.

\begin{subequations}
\label{pformIdentities}

\begin{eqnarray}
&& \star P_{(p)} = \frac{1}{(d-p)!}\left(\star P_{(p)} \right)_{a_1\cdots a_{d-p}}\dd x^{a_1}\wedge\cdots \wedge \dd x^{a_{d-p}}\,, \\
&& \quad\hbox{with} \quad  \left(\star P_{(p)} \right)_{a_1\cdots a_{d-p}}=\frac{1}{p!}\varepsilon_{a_1\cdots a_{d-p}}^{\phantom{a_1\cdots a_{d-p}}b_1\cdots b_p} P_{b_1\cdots b_p} \,, \nonumber
\end{eqnarray}

\begin{eqnarray}
&& P_{(p)} \wedge Q_{(q)} = \frac{1}{(p+q)!}\left( P_{(p)} \wedge Q_{(q)}\right)_{a_1\cdots a_{p+q}}\dd x^{a_1}\wedge\cdots \wedge \dd x^{a_{p+q}}\,, \\
&& \quad\hbox{with}\quad \left( P_{(p)} \wedge Q_{(q)} \right)_{a_1\cdots a_{p+q}}=\frac{(p+q)!}{p!q!}P_{[a_1\cdots a_p} Q_{a_{p+1}\cdots a_{p+q}]} \,, \nonumber
\end{eqnarray}

\begin{equation}
P_{(p)}\wedge Q_{(q)}= (-1)^{pq}Q_{(q)} \wedge P_{(p)} \,,
\end{equation}

\begin{equation}
A_{(p)}\wedge \star B_{(p)} = (-1)^p \star A_{(p)}\wedge B_{(p)} \,,
\end{equation}

\begin{equation}
\dd\left( P_{(p)}\wedge Q_{(q)}\right)=\dd P_{(p)}\wedge Q_{(q)}+(-1)^p P_{(p)}\wedge \dd Q_{(q)} \quad(\hbox{Graded Leibnitz rule})\,,
\end{equation}

\begin{eqnarray}
&& \ins_\xi P_{(p)} =\frac{1}{(p-1)!} ( \ins_\xi P_{(p)})_{a_1 \cdots a_{p-1}} \dd x^{a_1}\wedge\cdots \wedge \dd x^{a_{p-1}}\,, \\
&& \quad \hbox{with} \quad ( \ins_\xi P_{(p)})_{a_1 \cdots a_{p-1}} = \xi^c P_{c a_1 \cdots a_{p-1} }\,, \nonumber
\end{eqnarray}

\begin{equation}
\ins_\xi(P_{(p)}\wedge Q_{(q)})=(\ins_\xi P_{(p)})\wedge Q_{(q)}+(-1)^p P_{(p)}\wedge \ins_\xi Q_{(q)} \,,\nonumber
\end{equation}

\begin{equation}
\star\left( A_{a}^{\phantom{b}c_1\cdots c_{p-1}} B_{b c_1\cdots c_{p-1}}\xi^a \dd x^b\right)=(-1)^{p-1}(p-1)! \star B_{(p)}\wedge \ins_\xi A_{(p)}\,,
\end{equation}

\begin{equation}
\star\left( A_{a_1\cdots a_{p}}B^{a_1\cdots a_{p}}\xi_b \dd x^b\right)=(-1)^{p}p!  \ins_\xi \left(\star A_{(p)}\wedge B_{(p)}\right)\,. 
\end{equation}
\end{subequations}
\bibliography{refs}
\bibliographystyle{utphys}

\end{document}